\documentclass[galaxies,article,accept,pdftex,moreauthors]{Definitions/mdpi} 
\usepackage{graphicx}	
\usepackage{amsmath}	
\usepackage{amssymb}	
\usepackage{siunitx}
\firstpage{1} 
\pubvolume{1}
\issuenum{1}
\articlenumber{0}
\pubyear{2026}
\copyrightyear{2026}
\datereceived{ } 
\daterevised{ } 
\dateaccepted{ } 
\datepublished{ } 

\Title{Diffusion of PeV Cosmic Rays in the Turbulent and Multiphase Interstellar Medium}

\Author{Yue Hu $^{1}$\orcidA{}*}

\AuthorNames{Firstname Lastname, Firstname Lastname and Firstname Lastname}

\address{%
$^{1}$ \quad Institute for Advanced Study, 1 Einstein Drive, Princeton, NJ 08540, USA; yuehu@ias.edu
}

\corres{Correspondence: yuehu@ias.edu; NASA Hubble Fellow}

\abstract{Galactic cosmic rays (CRs) are a fundamental non-thermal component of the interstellar medium (ISM). Understanding the transport of superr-high-energy particles is essential for interpreting observations of Galactic PeVatrons. Classical diffusion models assuming a homogeneous and isothermal medium oversimplify the multiphase ISM. We utilize high-resolution three-dimensional magnetohydrodynamic simulations to self-consistently generate a multiphase ISM---comprising the warm (WNM), unstable (UNM), and cold neutral medium (CNM)---and investigate 1.5--15~PeV particle transport using a test-particle approach. We find that thermal phase transitions induce steep magnetic field strength gradients at phase boundaries, creating localized magnetic fluctuations that act as efficient sites for adiabatic mirror reflections and non-adiabatic pitch-angle scattering, strongly enhancing cross-field transport at these interfaces. However, because phase boundaries occupy only a small volume fraction and particles spend most of their trajectory in the weakly scattering WNM and UNM, the \textit{global} pitch-angle scattering coefficient in the multiphase ISM is smaller than in an equivalent isothermal medium. This locally strong scattering nevertheless drives both parallel and perpendicular spatial diffusion coefficients to $\sim 10^{30}$~cm$^2$~s$^{-1}$ at 1.5~PeV, with the perpendicular component exceeding its isothermal counterpart ($\sim 10^{28}$~cm$^2$~s$^{-1}$) by two orders of magnitude. Using a phase--phase diffusion matrix decomposition, we show that global CR transport is governed by the volume-filling, trans-Alfv\'enic WNM and UNM, where particles stream along stochastically wandering field lines. Cross-phase displacement correlations are universally positive, indicating cooperative transport between thermal phases. In contrast, the super-Alfv\'enic CNM acts as an efficient confinement that substantially suppresses local diffusion. }


\begin{document}

\section{Introduction}

\label{sec:intro}

Galactic cosmic rays (CRs) are a dynamically crucial component of the interstellar medium (ISM), significantly influencing its thermal evolution, ionization state, and overall energy balance \citep{2007ARNPS..57..285S,2008ApJ...674..258E,2008A&A...481...33J,10.5555/1593511,2009Natur.462..770V,2016ApJ...816L..19G,2017ApJ...834..208R,2021MNRAS.501.3640H,2022MNRAS.510.4952L,2023MNRAS.520.5126K,2025arXiv250118696H}. The energy spectrum of CRs spans a large range, adhering to a remarkably steady power law up to the so-called "knee" at approximately $3 \times 10^{15} \text{ eV}$ \citep{1990cup..book.....G,2007ARNPS..57..285S,2013FrPhy...8..748G,2017Sci...358..911A,2024PhRvL.132m1002C}. It is widely accepted that CRs below the knee are of Galactic origin, accelerated by energetic astrophysical engines such as supernova remnants, pulsar wind nebulae, and massive star clusters \citep{2013A&ARv..21...70B,2019IJMPD..2830022G,2022MNRAS.510.4952L,2025FrASS..1111076M}. However, identifying the exact locations of these extreme Galactic accelerators—dubbed "PeVatrons"—and understanding how PeV particles escape and propagate through the Galaxy has remained one of the most persistent challenges in high-energy astrophysics.

The experimental landscape was first transformed by the HAWC observatory, which identified extended TeV gamma-ray halos around middle-aged pulsars such as Geminga and Monogem \citep{2017Sci...358..911A}. These observations provided the first evidence of extremely suppressed, isotropic diffusion ($D \sim 10^{27} \text{cm}^2~\text{s}^{-1}$) for escaping leptonic particles in the local interstellar medium \citep{2017Sci...358..911A, 2019PhRvD.100l3015D}. A further breakthrough has been achieved by the Large High Altitude Air Shower Observatory (LHAASO), which has detected super-high-energy gamma-ray emission extending beyond $1 \text{ PeV}$ from multiple Galactic sources, including the Cygnus Cocoon, the Crab Nebula, and several SNRs \citep{2021Natur.594...33C,2021Sci...373..425L,2024PhRvL.132m1002C}. These super-high-energy photons serve as definitive proof for the existence of active Galactic PeVatrons, regardless of whether they originate from hadronic interactions between PeV protons and ambient gas, or from leptonic processes. Consequently, understanding the transport mechanics of both protons and electrons in the vicinity of these accelerators is essential to correctly interpret the spatial and spectral data, especially for explaining the highly symmetric and extended TeV--PeV gamma-ray halos that characterize many of these sources \citep{2021Natur.594...33C}.

The propagation of high-energy CRs in the Galaxy is classically modeled as a diffusion process, governed by the resonant pitch-angle scattering of particles off magnetic field fluctuations in the turbulent ISM \citep{1966ApJ...146..480J,1969ApJ...155..777J,2002PhRvL..89B1102Y,2002ApJ...578L.117Q,2008ApJ...673..942Y,2021ApJ...923...53L,2023MNRAS.525.4985K,2023JPlPh..89e1701L,2026MNRAS.545f2108E}. For CRs at $\text{PeV}$ energies, the self-generation of waves via the streaming instability is negligible due to the low number density of such energetic particles. Instead, their transport is almost entirely dictated by the extrinsic, pre-existing magnetohydrodynamic (MHD) turbulence injected by supernovae and galactic shear \citep{2002PhRvL..89B1102Y,2008ApJ...673..942Y,2013ApJ...779..140X,2022MNRAS.512.2111H,2023FrASS..1054760L,2025ApJ...994..142H}. However, characterizing this diffusion is highly non-trivial. While earlier studies have investigated GeV CR transport in multi-phase ISM \citep{2019A&A...622A.143C,2024ApJ...974...17H}, many previous theoretical and numerical studies on high-energy CR diffusion have predominantly assumed a simplified single-phase ISM \citep{2012JCAP...07..031G,2016MNRAS.457.3975S,2025ApJ...992...10K}. In reality, the Galactic ISM is highly inhomogeneous, existing in a multiphase state \citep{1999MNRAS.302..417S,2000ApJ...540..271V,2003ApJ...587..278W,2009ARA&A..47...27K,2011piim.book.....D,2018JCAP...08..049B,2019NatAs...3..776H,2024arXiv240714199H,2025ApJ...986...62H,vazquez2025interstellar}. It continuously transitions between the warm neutral medium (WNM, $T>5000$~K), the cold neutral medium (CNM, $T<200$~K), and the unstable neutral medium (UNM, $200<T<5000$~K ), dominantly driven by MHD turbulence \citep{2000ApJ...540..271V,2024arXiv240714199H,2025ApJ...986...62H}.

Because the different thermal phases exhibit drastically different densities, temperatures, and local magnetizations (i.e., varying sonic and Alfv\'enic Mach numbers, see \citep{2026arXiv260117255H}), the properties of the magnetic turbulence—and consequently, the fundamental CR diffusion coefficients (both parallel and perpendicular to the mean magnetic field)—must be phase-dependent \citep{2025arXiv251016125T,2026MNRAS.545f2108E}. A PeV CR injected from a local PeVatron will traverse a complex network of dense-cold and hot-diffuse structures. The assumption of a singular, volume-averaged diffusion coefficient breaks down in this realistic, highly structured ISM environment.

To overcome the limitations of isothermal models, we conduct a numerical investigation of PeV CR diffusion in a turbulent, magnetized, and multiphase ISM. We perform $2048^3$ high-resolution 3D MHD simulations to self-consistently generate a multiphase ISM exhibiting realistic WNM, UNM, and CNM distributions. By injecting $\gtrsim 1 \text{ PeV}$ test particles into this turbulent volume, we explicitly track their trajectories to measure the phase-dependent spatial diffusion coefficients. This approach allows us to directly connect the multi-phase turbulent properties to the transport of super-high-energy CRs, providing a crucial theoretical framework for interpreting the spatial morphology of the diffuse PeV gamma-ray emissions recently observed by LHAASO.

This paper is organized as follows. In \S~\ref{sec:method}, we detail the numerical methods used to generate the three-dimensional multiphase and isothermal ISM simulations, alongside our test-particle tracking framework. In \S~\ref{sec:results}, we present the statistical properties of the simulated turbulent media and provide a comprehensive analysis of the CR diffusion coefficients, highlighting the effects of scattering at phase-transition boundaries and phase-decomposed transport. Finally, we discuss the astrophysical implications of our findings and summarize our conclusions in \S~\ref{sec:discussion} and \S~\ref{sec:conclusion}.

\section{Numerical methods} 
\label{sec:method}
\subsection{MHD simulations of multi-phase ISM} 
In this work, we analyze a 3D simulation of the multiphase ISM utilizing the high-performance AthenaK code \citep{2024arXiv240916053S,2025ApJ...986...62H}. The simulation has been used in \citep{2026arXiv260117255H} for the study of turbulence in ISM and polarization dust emission. Here we briefly review the setup. The numerical model integrates the equations of ideal MHD within a periodic domain, expressed as:
\begin{equation}
\label{eq.mhd}
\begin{aligned}
&\frac{\partial\rho}{\partial t} +\nabla\cdot(\rho\boldsymbol{v})=0,\\
&\frac{\partial(\rho\boldsymbol{v})}{\partial t}+\nabla\cdot\left[\rho\boldsymbol{v}\boldsymbol{v}^T+\left(P+\frac{B^2}{8\pi}\right)\boldsymbol{I}-\frac{\boldsymbol{B}\boldsymbol{B}^T}{4\pi}\right] = \boldsymbol{f},\\
&\frac{\partial\boldsymbol{B}}{\partial t} - \nabla\times(\boldsymbol{v}\times\boldsymbol{B})=0,\\
&\nabla \cdot\boldsymbol{B}=0,
\end{aligned}
\end{equation}
alongside the total energy equation:
\begin{equation}
\label{eq.energy}
\begin{aligned}
&\frac{\partial E}{\partial t} + \nabla \cdot \left[\boldsymbol{v}\left(E + P + \frac{B^2}{8\pi}\right) - \frac{\boldsymbol{B}(\boldsymbol{B}\cdot\boldsymbol{v})}{4\pi}\right] = \Gamma - \Lambda + \boldsymbol{f}\cdot\boldsymbol{v},
\end{aligned}
\end{equation}
Here, $\rho$ and $\boldsymbol{v}$ are gas mass density and gas velocity, respectively. $E$ is the gas energy density and $\boldsymbol{B}$ is magnetic field. $P$ denotes the thermal gas pressure, $\mathbf{I}$ represents the identity tensor, and $\boldsymbol{f}$ introduces a stochastic continuous forcing mechanism to sustain the turbulent flow.

To model the thermodynamic phase transitions, we incorporate an atomic line cooling function, $\Lambda$, parameterized following the fits by \citep{2002ApJ...564L..97K}:
\begin{equation}
\label{eq:cooling}
\begin{aligned}
\Lambda = \left(\frac{\rho}{m_{\rm H}}\right)^2 & \left[ 2 \times 10^{-19}\exp\left(\frac{-1.148 \times 10^5}{T+1000}\right) \right. \\
& \left. + 2.8 \times 10^{-28}\sqrt{T}\exp\left(\frac{-92}{T}\right) \right],
\end{aligned}
\end{equation}
where $m_{\rm H}$ stands for the hydrogen mass, with $\Lambda$ evaluated in units of ${\rm erg~s^{-1}~cm^{-3}}$. Conversely, a uniform volumetric heating rate, $\Gamma$, is applied as:
\begin{equation}
\label{eq:heating}
\Gamma = \left(\frac{2\rho}{m_{\rm H}}\right) \times 10^{-26},
\end{equation}
which is in units of ${\rm erg~s^{-1}~cm^{-3}}$. Thermal conduction is explicitly omitted from our physical model; its associated dissipation scale ($\sim0.01$ pc) remains negligibly small relative to the ISM scales of interest \citep{2000ApJ...540..271V,2024arXiv240714199H,2025ApJ...986...62H}.

The simulation initializes with a homogeneous gas density of $n = 1~{\rm cm^{-3}}$—yielding an average column density of $N_{\rm H} = 3 \times 10^{20}~{\rm cm^{-2}}$—permeated by a uniform, $z$-directed mean magnetic field. We set this initial background field strength to $B \approx$ 3 \SI{}{\micro G} to align with standard Zeeman splitting measurements in the Galaxy \citep{2012ARA&A..50...29C}. The initial turbulent velocity dispersion is prescribed as $\sigma_v \approx 10~{\rm km~s^{-1}}$, a value anchored in Larson's empirical laws and typical conditions observed within young star-forming regions \citep{1981MNRAS.194..809L,2022ApJ...934....7H}.

We perform the simulation within a cubic volume spanning $100~{\rm pc}$ on each side, mapped onto a high-resolution $2048^3$ Cartesian grid. This spatial discretization effectively restricts numerical dissipation to a scale of roughly 10 computational cells. Kinetic energy is injected via purely solenoidal turbulent driving at a scale of $L_{\rm inj} = 50~{\rm pc}$, equivalent to a peak wavenumber of $k = 2\pi/L_{\rm inj} = 0.125~{\rm pc^{-1}}$. Finally, the entire system is integrated over 100 Myr, ensuring that the MHD turbulence has fully reached a statistically stationary state. Details of turbulence properties are given in \cite{2026arXiv260117255H}.

\subsection{3D isothermal MHD turbulence simulation}
For comparison, we also conduct a 3D isothermal MHD turbulence control simulation using the AthenaK code \citep{2024arXiv240916053S,2024MNRAS.527.3945H}. This run solves the ideal MHD equations (see Eq.~\ref{eq.mhd}) subject to periodic boundary conditions. To isolate the kinematic effects of turbulence from the thermodynamic phase transitions, we replace the explicit heating and cooling source terms with a simple isothermal equation of state, $P = \rho c_s^2$, where $\rho$ is the local gas mass density and $c_s$ is the constant isothermal sound speed. The system is initialized with a uniform density field and a uniform background magnetic field aligned along the $z$-axis.

Turbulence is continuously driven by injecting kinetic energy via a purely solenoidal forcing field that peaks at wavenumber $k = 2$. The system is naturally evolved for at least one sound-crossing time until the turbulence becomes fully developed, yielding a steady-state Kolmogorov-like cascade. The computational domain is discretized onto a uniform $1200^3$ grid. To provide a baseline for comparison, the driving amplitude and initial magnetic field strength are tuned to yield a global sonic Mach number of $M_s \sim 1$ and an Alfv\'enic Mach number of $M_A \sim 1.5$, closely matching the volume-averaged turbulent properties of the multi-phase simulation. Nevertheless, a same resolution match is more ideal.

\subsection{Test particle ejection}
To investigate the transport of PeV CR within the multiphase ISM, we employ a test-particle tracking approach \citep{2013ApJ...779..140X}. Because the velocity of cosmic rays is orders of magnitude larger than the characteristic dynamical velocities of the ISM (e.g., the Alfv\'en and sound speeds), the turbulent magnetic field is effectively stationary over the particle scattering timescales. Therefore, we extract 3D magnetic field snapshots from the AthenaK simulations once the turbulence has reached a statistically steady state. We integrate the collisionless equation of motion governed purely by the magnetic Lorentz force:
\begin{equation}
\frac{d\mathbf{u}}{dt} = \frac{q}{m\gamma c} (\mathbf{u} \times \mathbf{B}),
\end{equation}
where the particle speed $|\mathbf{u}|$ remains strictly conserved. To ensure a statistically representative sampling of the multiphase environment, we randomly inject test particles across the entire $100$~pc computational domain and trace the trajectory for 1,000 gyro periods. 
The initial particle velocity vectors are distributed isotropically, with the pitch-angle cosine ($\mu = \cos\theta$) sampled uniformly from $[-1, 1]$. We also perform controlled runs with mono-pitch-angle injections to isolate specific scattering regimes.

We set the particle Larmor radius to $r_L = 10 - 100$ grid cells. Given our domain size ($L = 100$~pc) and resolution ($2048^3$), a single grid cell corresponds to $\approx 0.05$~pc, yielding a physical Larmor radius of $r_L \approx 0.5 - 5$~pc. In a mean Galactic magnetic field of $B \approx$ 3 \SI{}{\micro G}, this gyroradius precisely corresponds to a CR energy of $E \approx 1.5 - 15$~PeV, directly aligning with the energy regime probed by recent LHAASO observations.

\subsection{Total Diffusion Coefficients}\label{sec:total_diff}
Before decomposing by phase, we first measure the \emph{global} spatial and pitch-angle diffusion coefficients by tracking particle-averaged mean-squared displacements over the full trajectory.

\subsubsection{Spatial Diffusion via Mean-Squared Displacement}\label{sec:msd_total}
We trace 1000 test particles through a static snapshot, which is statistically sufficient as shown in \citep{2013ApJ...779..140X}. For each particle $i$, the net displacement at a common reference time $t$ is $\Delta\boldsymbol{r}_{i} = \boldsymbol{r}_{i}(t) - \boldsymbol{r}_{i}(0)$.  The parallel ($\parallel$) and perpendicular ($\perp$) directions are defined with respect to the mean magnetic field $\hat{\boldsymbol{B}_0}$.  For each displacement vector $\Delta\boldsymbol{r}_i$, the parallel and perpendicular components are obtained from projecting onto the mean-field direction:
\begin{align}
  \Delta r_{i,\parallel} &= \Delta\boldsymbol{r}_{i}\cdot\hat{\boldsymbol{B}_0},
    \label{eq:dr_para_global}\\
  |\Delta\boldsymbol{r}_{i,\perp}|^{2} &=
    |\Delta\boldsymbol{r}_{i}|^{2} - (\Delta r_{i,\parallel})^{2}.
    \label{eq:dr_perp_global}
\end{align}

The ensemble-averaged diffusion coefficients are then
\begin{align}
  D_{\parallel} &=
    \frac{1}{N}\sum_{i=1}^{N}
    \frac{(\Delta r_{i,\parallel})^{2}}{2\,t},
    \label{eq:Dpara_total}\\[4pt]
  D_{\perp} &=
    \frac{1}{N}\sum_{i=1}^{N}
    \frac{|\Delta\boldsymbol{r}_{i,\perp}|^{2}}{4\,t},
    \label{eq:Dperp_total}
\end{align}
where the denominators $2t$ and $4t$ correspond to diffusion in one and two spatial dimensions, respectively.  

\subsubsection{Pitch-Angle Diffusion}\label{sec:Dmumu}

The pitch-angle cosine of particle $i$ with respect to the local magnetic field is $\mu_{i}(t)$.  Its diffusion coefficient is
\begin{equation}\label{eq:Dmumu}
  D_{\mu\mu} =
    \frac{1}{N}\sum_{i=1}^{N}
    \frac{[\mu_{i}(t)-\mu_{i}(0)]^{2}}{2\,t},
\end{equation}
which represents the rate of pitch-angle scattering. 

\subsection{Phase-Decomposed Diffusion Analysis}\
\label{sec:phase_decomp}
In a multiphase ISM, CR transport is governed by the statistical properties of magnetic turbulence, which vary
substantially among thermally distinct phases.  We develop a formalism that \emph{exactly} decomposes the total spatial diffusion coefficient into a
matrix of phase--phase contributions, revealing the role of each phase.   

\subsubsection{Displacement Decomposition}
\label{sec:displacement}
The total displacement of a particle over the interval $[0,\,t]$ is:
\begin{equation}\label{eq:decomp}
  \Delta\boldsymbol{r}(t) =
  \sum_{\alpha=1}^{N_{p}}\Delta\boldsymbol{r}_{\alpha}(t),
\end{equation}
where $\Delta\boldsymbol{r}_{\alpha}(t)$ accumulates displacement \emph{only} during intervals when the particle is in phase $\alpha$; It is zero during all other intervals. 

The ensemble-averaged mean-squared displacement (MSD) is obtained by squaring Eq.~(\ref{eq:decomp}) and averaging over particles:
\begin{equation}\label{eq:msd_expand}
  \bigl\langle|\Delta\boldsymbol{r}|^{2}\bigr\rangle =
  \biggl\langle\biggl|\sum_{\alpha}\Delta\boldsymbol{r}_{\alpha}
  \biggr|^{2}\biggr\rangle =
  \sum_{\alpha=1}^{N_{p}}\sum_{\beta=1}^{N_{p}}
  \bigl\langle
    \Delta\boldsymbol{r}_{\alpha}\cdot\Delta\boldsymbol{r}_{\beta}
  \bigr\rangle.
\end{equation}
This identity decomposes the total MSD into a symmetric $N_{p}\times N_{p}$ matrix of cross-correlation terms.

\begin{figure*}
\centering
\includegraphics[width=0.99\linewidth]{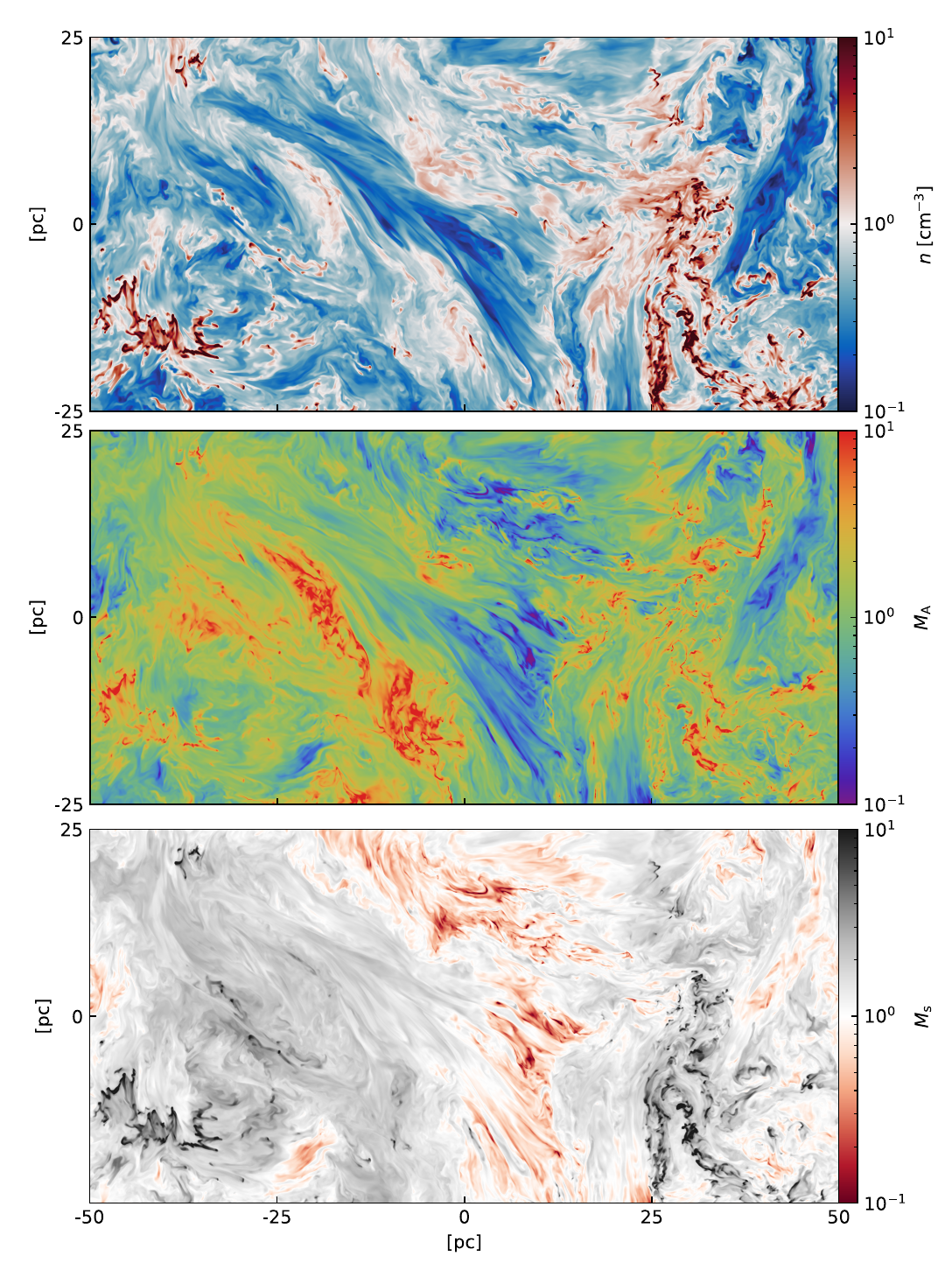}
        \caption{Two-dimensional spatial slices from the $2048^3$ multiphase ISM simulation, illustrating the gas number density ($n$, top), the local Alfv\'enic Mach number ($M_A = v/v_A$, middle), and the sonic Mach number ($M_s = v/c_s$, bottom), where $v$ is the local turbulent velocity, $v_A$ is the Alfv\'en speed, and $c_s$ is the sound speed. For visualization purposes, the vertical axis is truncated to display a $100 \times 50$ pc sub-region, representing half of the full simulation slice. The spatial correlation between the panels highlights the multiphase nature of the medium, showing how dense, clumpy density structures coincide with regions of supersonic and super-Alfv\'enic turbulence (high $M_s$ and high $M_A$).}
    \label{fig:maps}
\end{figure*}

\subsubsection{Phase--phase diffusion coefficient matrix.}
Projecting onto the parallel and perpendicular directions, we define the \emph{phase--phase diffusion coefficient matrix}:
\begin{align}
  D_{\alpha\beta}^{\parallel}(t) &\equiv
    \frac{\bigl\langle
      \Delta r_{\alpha,\parallel}(t)\,
      \Delta r_{\beta,\parallel}(t)
    \bigr\rangle}{2\,t},
    \label{eq:Dpar}\\[6pt]
  D_{\alpha\beta}^{\perp}(t) &\equiv
    \frac{\bigl\langle
      \Delta\boldsymbol{r}_{\alpha,\perp}(t)\cdot
      \Delta\boldsymbol{r}_{\beta,\perp}(t)
    \bigr\rangle}{4\,t},
    \label{eq:Dperp}
\end{align}
where $\Delta r_{\alpha,\parallel} = \Delta\boldsymbol{r}_{\alpha}\cdot\hat{\boldsymbol{B}_0}$ and $\Delta\boldsymbol{r}_{\alpha,\perp} =
\Delta\boldsymbol{r}_{\alpha} - \Delta r_{\alpha,\parallel}\,\hat{\boldsymbol{B}_0}$.  The factor of $4$ in Eq.~(\ref{eq:Dperp}) accounts for diffusion in two transverse dimensions.  In practice, $\langle\cdots\rangle$ denotes an average over all particles. From Eqs.~(\ref{eq:msd_expand})--(\ref{eq:Dperp}), the total diffusion coefficients satisfy:
\begin{equation}\label{eq:sumrule}
  D_{\parallel}(t) =
    \sum_{\alpha,\beta}D_{\alpha\beta}^{\parallel}(t),
  \qquad
  D_{\perp}(t) =
    \sum_{\alpha,\beta}D_{\alpha\beta}^{\perp}(t).
\end{equation}
This sum rule holds \emph{exactly} at every time $t$ and serves as an internal consistency check of the numerical implementation. We track the time evolution of $D_{\parallel}$ and $D_{\perp}$ and take their saturated values at the late-time plateau as the final diffusion coefficients.

The diagonal elements $D_{\alpha\alpha}$ quantify the contribution of scattering \emph{within} phase $\alpha$ to the total diffusion. A phase that occupies a large volume fraction and efficiently scatters particles will produce a large $D_{\alpha\alpha}$. The off-diagonal elements $D_{\alpha\beta}$ ($\alpha\neq\beta$) encode the \emph{cross-correlation} between displacements accumulated in different phases.  Their sign carries physical meaning:
\begin{itemize}
  \item $D_{\alpha\beta}>0$: the displacements in phases $\alpha$ and $\beta$ are positively correlated---particles tend to drift in the same direction in both phases, cooperatively enhancing diffusion.
  \item $D_{\alpha\beta}<0$: the displacements are anti-correlated---the drift acquired in one phase is partially reversed in the other, suppressing diffusion.
  \item $D_{\alpha\beta}\approx 0$: the displacements in the two phases are statistically uncorrelated.
\end{itemize}

\begin{figure*}
\centering
\includegraphics[width=0.99\linewidth]{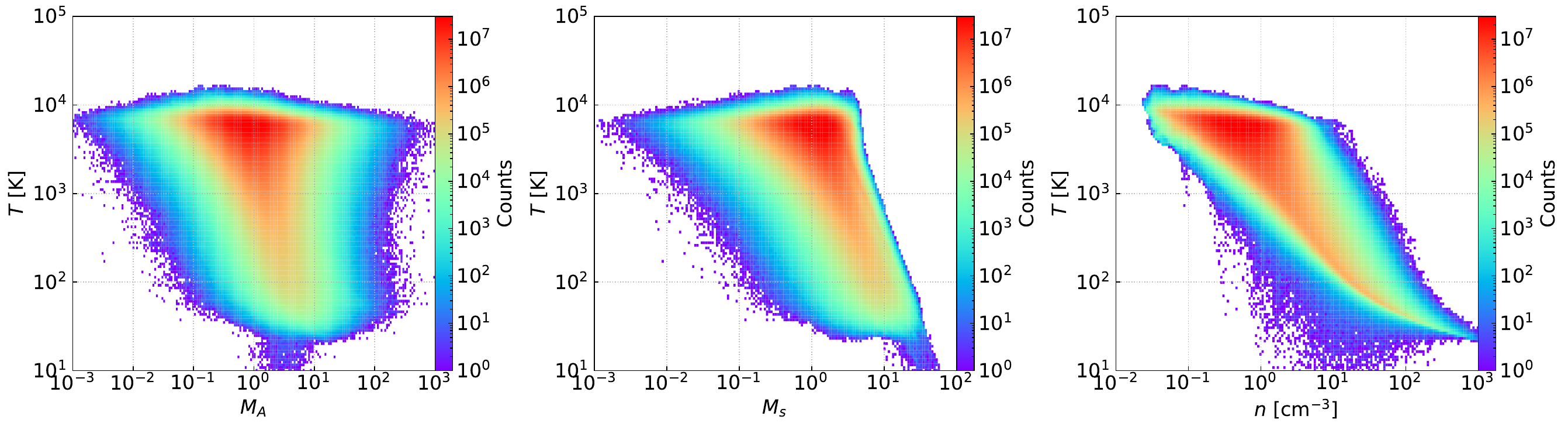}
        \caption{Two-dimensional histograms showing the distribution of gas properties in the simulated multiphase interstellar medium. From left to right, the panels display the gas temperature ($T$) as a function of the local Alfv\'enic Mach number ($M_A$), sonic Mach number ($M_s$), and gas number density ($n$). The color scale indicates the number of computational cells (Counts) within each bin on a logarithmic scale. 
        }
    \label{fig:2Dhist}
\end{figure*}

\section{Results}
\label{sec:results}
\subsection{Properties of the multi-phase ISM}
The presence of MHD turbulence naturally drives the interstellar gas into a highly inhomogeneous, multiphase state \citep{2000ApJ...540..271V,2024arXiv240714199H,2025ApJ...986...62H}. Fig.~\ref{fig:maps} displays two-dimensional spatial slices from the $2048^3$ simulation, showing the complex morphological structures generated by these dynamics. The top panel reveals the highly structured gas number density ($n$), characterized by a web of dense and cold clumps embedded within a voluminous, hot, and diffuse background. The interplay between turbulent mixing and magnetic fields—both of which dynamically support the unstable transitional gas—results in global mass fractions of 21.79\% for the CNM, 33.93\% for the WNM, and 44.28\% for the UNM.

The corresponding $M_s$ slice (Fig.~\ref{fig:maps}, bottom panel) shows a spatial correspondence with the density field. Because the sound speed drops precipitously in the cold gas, the dense filaments systematically coincide with regions of intensely supersonic turbulence. In contrast, the local $M_A$ slice (middle panel) exhibits a highly complex distribution. The magnetic field experiences dynamo, stretching, and compression, leading to a highly variable $M_A$ distribution. 

To quantitatively characterize these distinct thermal phases, Fig.~\ref{fig:2Dhist} presents 2D histograms of the gas properties across the entire simulated volume. From left to right, the panels display the gas temperature ($T$) as a function of the local $M_A$, the $M_s$, and the gas number density ($n$). The rightmost panel clearly illustrates the thermal phase diagram of the WNM (high $T$, low $n$) and the CNM (low $T$, high $n$), separated by the transitional UNM. Specifically, the WNM is predominantly distributed around $n \sim 0.1 - 1~\rm{cm^{-3}}$ at $T \sim 10^4~\rm{K}$, while the CNM clusters at $n \sim 10 - 100~\rm{cm^{-3}}$ with temperatures generally below $100~\rm{K}$.

\begin{figure}
\centering
\includegraphics[width=0.5\linewidth]{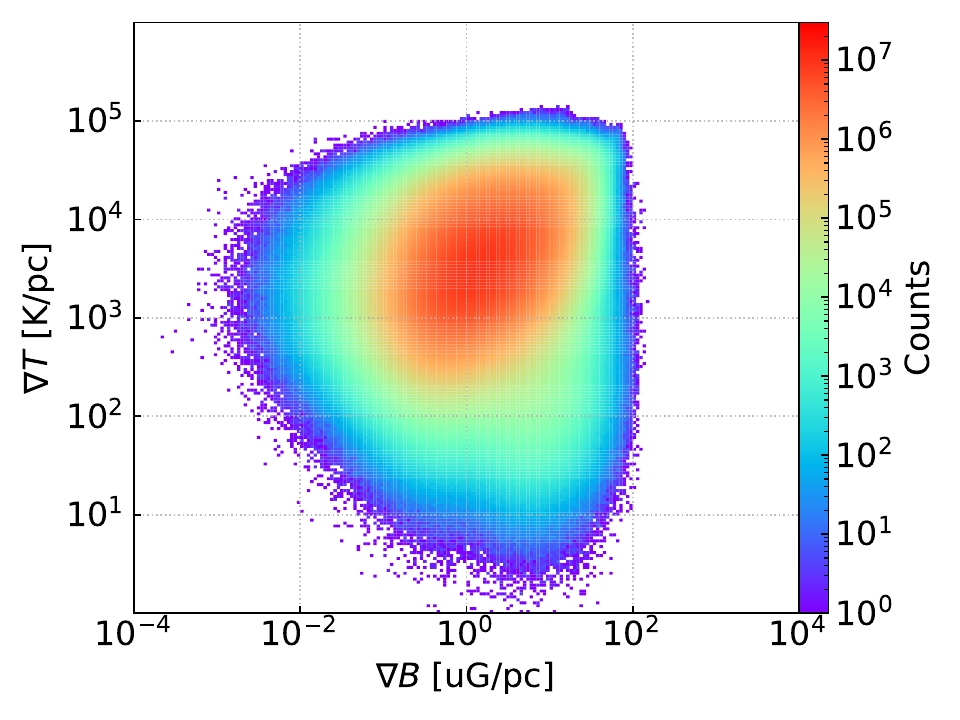}
        \caption{Two-dimensional histograms showing the spatial gradient of gas temperature and the gradient of magnetic field strength. The color scale indicates the number of computational cells (Counts) within each bin on a logarithmic scale.}
    \label{fig:gradT_gradB}
\end{figure}

\begin{figure*}[p]
\centering
\includegraphics[width=0.99\linewidth]{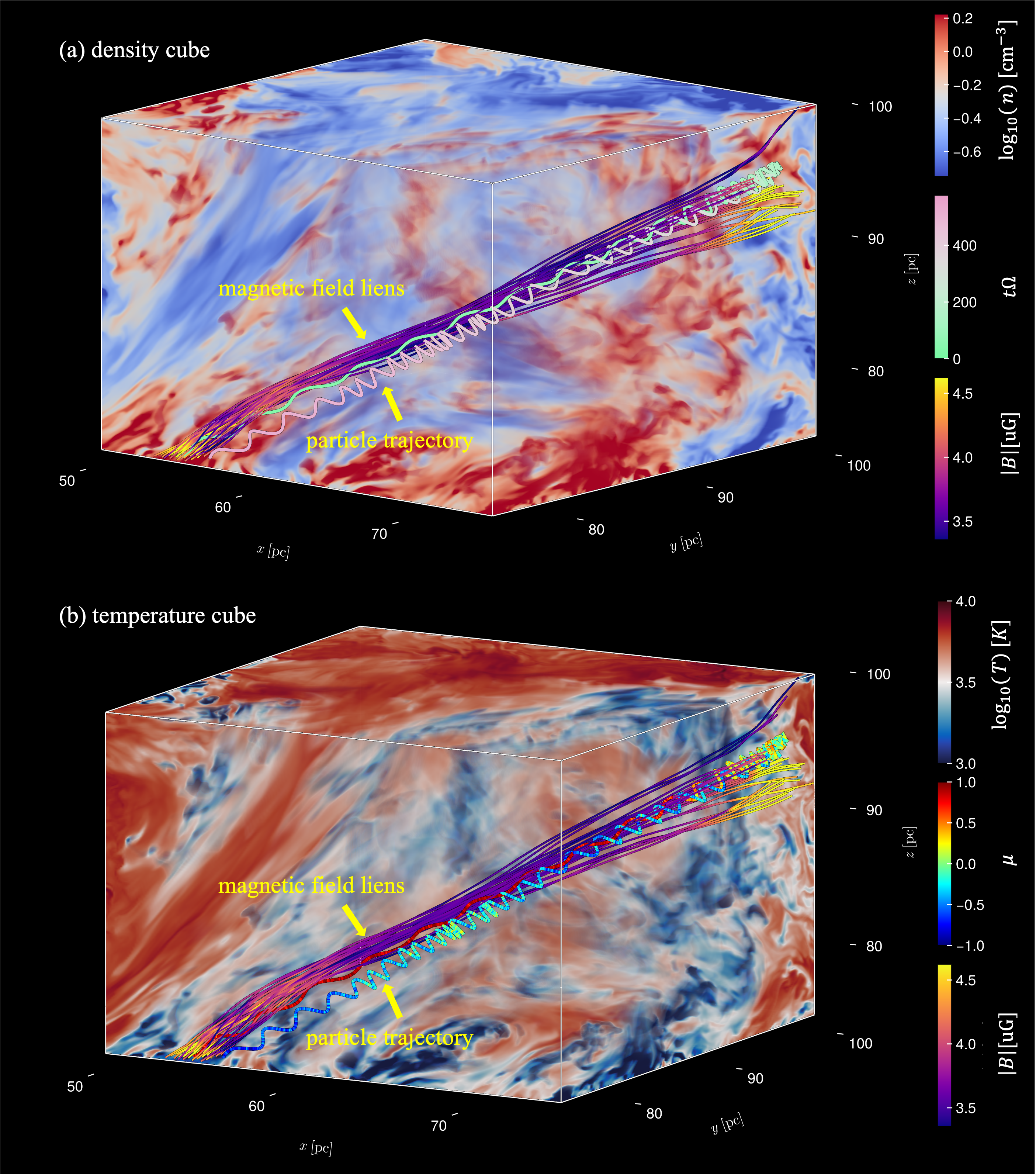}
        \caption{Numerical illustration of particle transport through a multiphase ISM, consisting of the WNM, the UNM, and the CNM. Panel (a) displays a 3D volume rendering of the gas number density ($\log_{10}(n) \: [\text{cm}^{-3}]$), highlighting the complex distribution of dense structures and diffuse cavities. Panel (b) illustrates the corresponding gas temperature ($\log_{10}(T)~[\text{K}]$). In both panels, a bundle of local magnetic field lines is overlaid, color-coded by the magnetic field strength ($|B|$ [\SI{}{\micro G}]). A representative trajectory of a charged particle is superimposed to demonstrate its helical gyromotion along the fluctuating magnetic field. The particle's path is color-coded by the normalized time ($t\Omega$, where $\Omega$ is the gyrofrequency) in panel (a), and by the pitch-angle cosine ($\mu$) in panel (b), tracking the evolution of the particle's scattering and pitch-angle variations as it propagates through the turbulent ISM.}
    \label{fig:illustration}
\end{figure*}

Furthermore, the left and middle panels of Fig.~\ref{fig:2Dhist} show that the turbulent properties and intrinsic magnetization vary dynamically across these different thermal phases. The middle panel reveals a strong temperature dependence for the sonic Mach number: the hot, diffuse WNM is predominantly subsonic to transonic ($M_s \lesssim 1$), whereas the cold, dense CNM is highly supersonic, with $M_s$ extending up to $\sim 10 - 20$ due to its significantly lower sound speed. Similarly, the left panel indicates a clear shift in the $M_A$ across the phases. While the WNM is broadly trans-Alfv\'enic ($M_A \sim 1$), the CNM exhibits a wider distribution that extends deep into the super-Alfv\'enic regime ($M_A \gtrsim 1$, which means the turbulent kinetic energy is larger than the magnetic field energy). The CNM, however,  has a plasma beta, suggesting thermal pressure is weaker than magnetic field pressure. A more detailed analysis of its turbulence properties and corresponding physical explanations is given in \cite{2026arXiv260117255H}. 

\subsection{Phase transition induces large magnetic field fluctuations}
The nonlinear thermal phase transitions profoundly alter not only the density distribution but also the magnetic field topology. Fig.~\ref{fig:gradT_gradB} shows a two-dimensional histogram correlating the spatial gradient of the gas temperature ($|\nabla T|$) with the gradient of the magnetic field strength ($|\nabla B|$).  The distribution reveals a clear positive correlation, with the peak cell counts heavily clustered around sharp temperature gradients of $|\nabla T| \sim 10^3-10^4~{\rm K~pc^{-1}}$ and correspondingly strong magnetic gradients of $|\nabla B| \sim 1-10$ \SI{}{\micro G}${\rm~pc^{-1}}$. Furthermore, the distribution tail extends to strong magnetic field gradients approaching $|\nabla B| \sim 10^2$ \SI{}{\micro G}${\rm~pc^{-1}}$ precisely in regions where the temperature gradient remains highly elevated. This demonstrates that regions experiencing steep temperature variations—namely, the narrow boundaries between distinct thermal phases—simultaneously host intense magnetic fluctuations. These localized, extreme magnetic gradients naturally create magnetic bottlenecks, or "mirrors," throughout the multiphase ISM.

\subsection{Strong CR scattering at phase transition boundaries}
Fig.~\ref{fig:illustration} provides a three-dimensional visualization of CR transport within the turbulent and multiphase ISM. As the particle initially propagates along the mean magnetic field, its pitch-angle cosine ($\mu$, indicated by the color scale in panel b) undergoes a gradual, continuous evolution driven by pitch-angle scattering from turbulent fluctuations. As the particle penetrates deeper into the domain, it encounters a localized region of significantly enhanced magnetic field strength, visibly highlighted by the yellow segments of the magnetic field lines. At this magnetic bottleneck, the particle is dynamically reflected via the magnetic mirror effect, abruptly reversing its propagation direction. This reflection is clearly indicated by the sharp transition in $\mu$ from a positive to a negative value. 
Crucially, this reflection event promotes efficient cross-field transport through two channels. Where the magnetic field gradient is sufficiently gentle relative to the particle gyroradius ($|\nabla B|\, r_L / B \lesssim 1$), the reflection is adiabatic, and the perpendicular displacement arises from the lateral wandering of the field line between mirror bounces, consistent with the mirror diffusion framework \citep{2021ApJ...923...53L}. At the sharpest phase boundaries, however, the magnetic field strength varies on spatial scales comparable to $r_L$, violating the first adiabatic invariant $\mu_\mathrm{mag} = m u_\perp^2/2B$ and producing a direct perpendicular displacement of the particle's guiding center during the interaction. In both cases, the net effect is an irreversible decoupling of the CR from its original field line, providing a highly efficient mechanism for driving the perpendicular spatial diffusion of CRs in the multiphase environment.


\begin{figure*}
    \centering
    \subfloat[Multiphase ISM case. The top row plots the pitch-angle scattering rate against the rate of change of the local gas temperature, $|{\rm d}T/{\rm d}(t\Omega)|$, while the bottom row plots it against the rate of change of the magnetic field strength, $|{\rm d}B/{\rm d}(t\Omega)|$.  Columns correspond to $E = 1.5$, $7.5$, and $15$ PeV.\label{fig:T_B_mu}]{%
        \includegraphics[width=\linewidth]{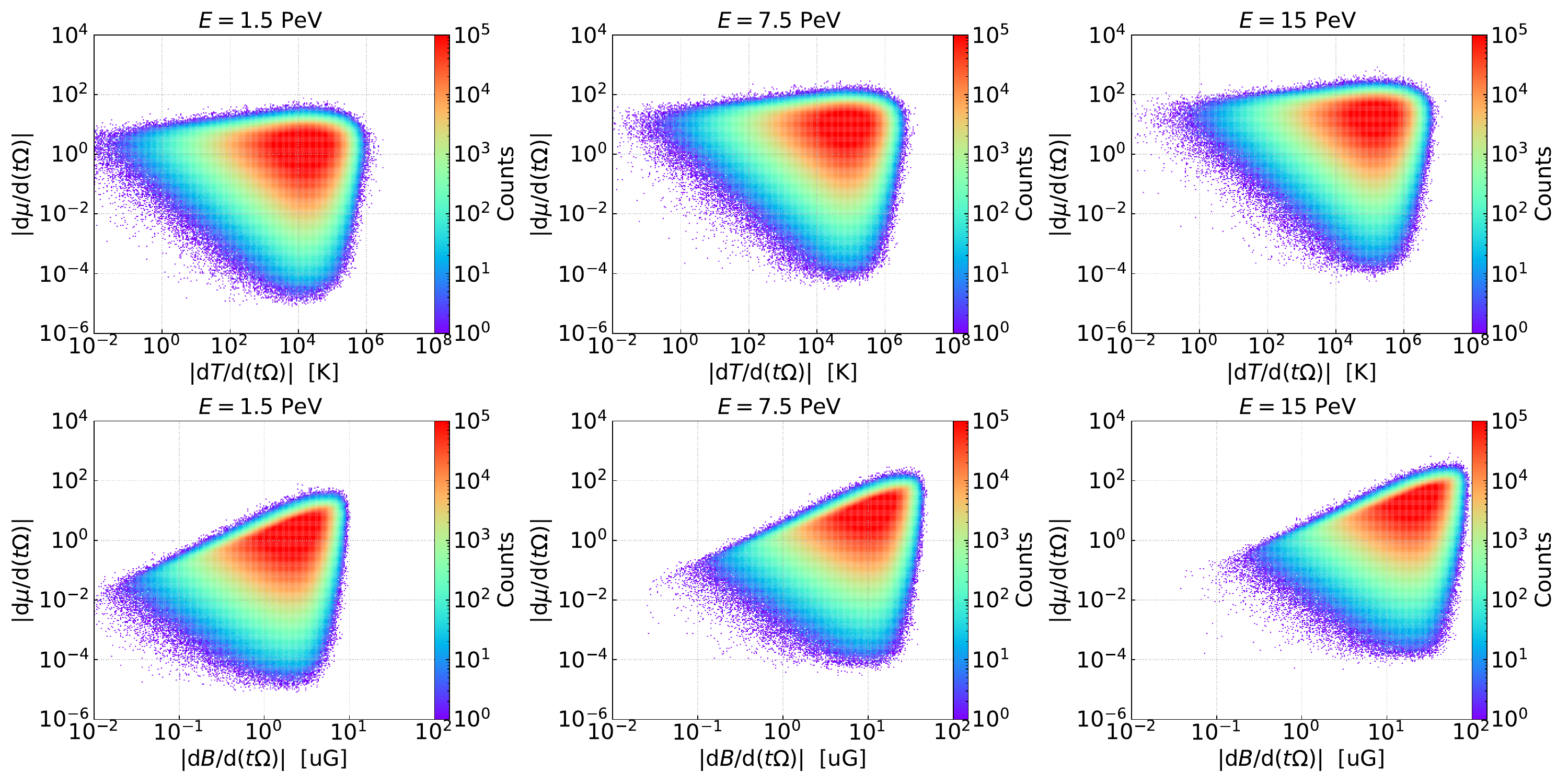}%
    }
    \\ 
    \subfloat[Isothermal ISM case. The plots show $|{\rm d}\mu/{\rm d}(t\Omega)|$ versus $|{\rm d}B/{\rm d}(t\Omega)|$, corresponding to $E = 2.5$, $7.5$, and $15$ PeV.\label{fig:B_mu}]{%
        \includegraphics[width=\linewidth]{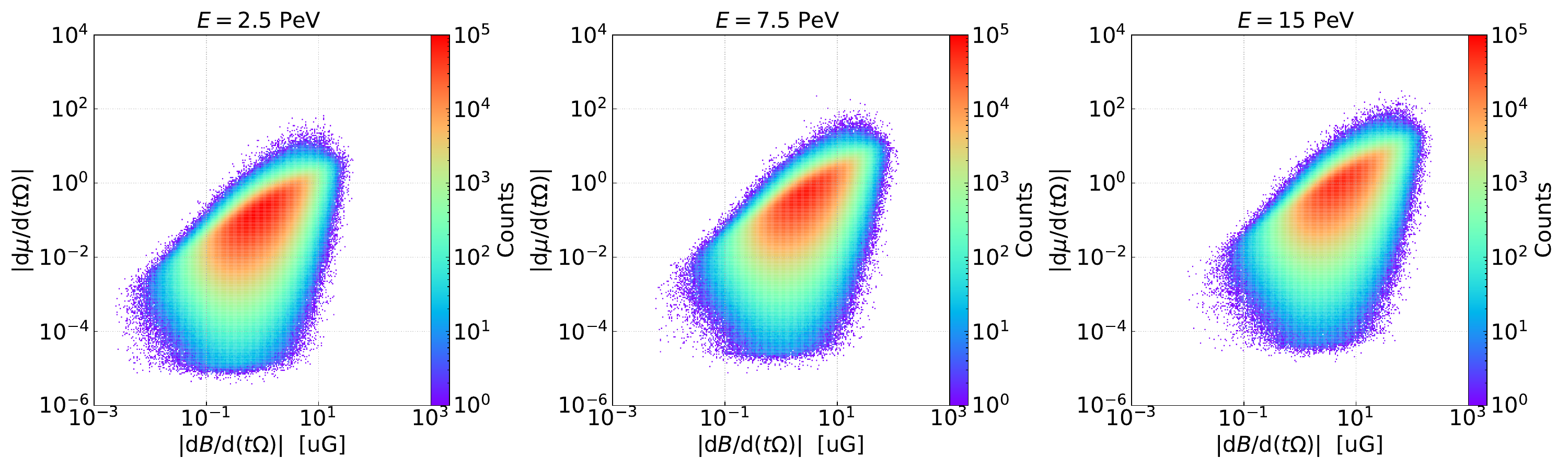}%
    }
    \caption{Two-dimensional histograms illustrating the scattering dynamics of CRs propagating through the multiphase ISM (a) or isothermal medium (b). The vertical axes in all panels display the absolute rate of change of the pitch-angle cosine, $|{\rm d}\mu/{\rm d}(t\Omega)|$, normalized by the particle gyrofrequency ($\Omega$). The color scale indicates the logarithmic cell counts for each bin.}
    \label{fig:T_B_mu_combined}
\end{figure*}



Fig.~\ref{fig:T_B_mu} shows the relationship between phase-transition-induced magnetic fluctuations and pitch-angle scattering. The panels display the absolute rate of change of the pitch-angle cosine, $|{\rm d}\mu/{\rm d}(t\Omega)|$, normalized by the particle gyrofrequency ($\Omega$), as a function of the corresponding rates of change in the magnetic field strength, $|{\rm d}B/{\rm d}(t\Omega)|$, and the gas temperature, $|{\rm d}T/{\rm d}(t\Omega)|$. Because the cosmic rays propagate at the speed of light ($v \approx c$), these temporal derivatives are evaluated directly along the particle trajectories, effectively capturing the spatial gradients encountered during transport. The two-dimensional histograms exhibit a clear positive correlation between $|{\rm d}\mu/{\rm d}(t\Omega)|$ and both $|{\rm d}B/{\rm d}(t\Omega)|$ and $|{\rm d}T/{\rm d}(t\Omega)|$. Notably, the bottom panels reveal an apparent diagonal trend: the most frequent, large-angle scattering events (peaking at $|{\rm d}\mu/{\rm d}(t\Omega)| \sim 10^{-1}-10^1$) coincide exactly with sharp, localized magnetic field jumps of $|{\rm d}B/{\rm d}(t\Omega)| \sim 1-10$ \SI{}{\micro G}. Simultaneously, the top panels show these intense scattering clusters broadly around extreme temperature variations of $|{\rm d}T/{\rm d}(t\Omega)| \sim 10^3-10^6~{\rm K}$. This alignment demonstrates that instantaneous, large-angle scattering events—characterized by abrupt changes in the pitch angle—are directly driven by particles traversing the extreme temperature and magnetic field gradients native to multiphase transition boundaries.

It is important to note that a positive correlation between $|{\rm d}\mu/{\rm d}(t\Omega)|$ and $|{\rm d}B/{\rm d}(t\Omega)|$ naturally exists in both the multiphase and isothermal media (see Fig.~\ref{fig:B_mu}). However, the nature of the scattering differs. In the isothermal medium, the scattering (the red peak in Fig.~\ref{fig:T_B_mu}) occurs at moderate magnetic gradients ($|{\rm d}B/{\rm d}(t\Omega)| \sim 10^{-1}~\SI{}{\micro G}$) and results in relatively mild pitch-angle diffusion ($|{\rm d}\mu/{\rm d}(t\Omega)| \sim 10^{-2}$). This reflects particles interacting with gradual, continuous turbulent cascades. In contrast, the multiphase medium (Fig.~\ref{fig:T_B_mu}, bottom row) shifts the core scattering distribution to higher extremes. The peak scattering is driven by localized magnetic gradients that are an order of magnitude steeper ($|{\rm d}B/{\rm d}(t\Omega)| \sim 1-\SI{10}{\micro G}$), causing instantaneously violent pitch-angle changes ($|{\rm d}\mu/{\rm d}(t\Omega)| \sim 10^0$). This difference suggests that while isothermal scattering is dominated by gyroresonance with turbulence cascades, multiphase scattering is heavily dominated by abrupt, hard reflections at the extreme magnetic fluctuations formed by thermal phase boundaries.

\begin{figure}
\centering
\includegraphics[width=0.5\linewidth]{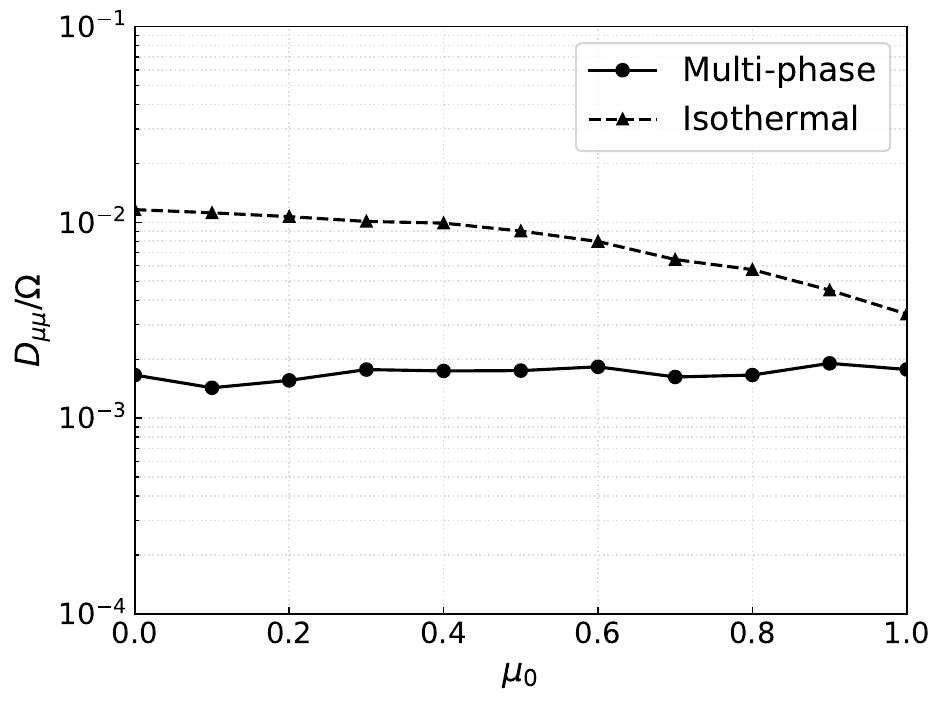}
        \caption{ Normalized pitch-angle diffusion coefficient, $D_{\mu\mu}/\Omega$, as a function of the initial pitch-angle cosine, $\mu_0$. The solid and dashed lines represent the results from the multi-phase and isothermal ISM simulations, respectively. Particle energy is $\sim2.5$ PeV.}
    \label{fig:Dmu}
\end{figure}

Fig.~\ref{fig:Dmu} further illustrates the normalized pitch-angle diffusion coefficient ($D_{\mu\mu}/\Omega$) as a function of the initial pitch-angle cosine ($\mu_0$) at 2.5 PeV. Across the entire range from $\mu_0 = 0$ to 1, the pitch-angle diffusion coefficient in the multiphase case is around $D_{\mu\mu}/\Omega \sim 2 \times 10^{-3}$. The corresponding coefficient in the isothermal medium is $D_{\mu\mu}/\Omega \sim 1 \times 10^{-2}$ at $\mu_0=0$ and decreases to $3 \times 10^{-3}$. This difference highlights the fundamentally intermittent nature of CR transport in the multiphase ISM. In the isothermal medium, magnetic fluctuations are volume-filling, subjecting particles to continuous, moderate scattering everywhere. In contrast, intense scattering in the multiphase ISM is heavily localized to the narrow phase-transition boundaries and the small-volume CNM clumps. For most of their trajectory, particles propagate with minimal scattering through the sub-Alfv\'enic/trans-Alfv\'enic WNM and UNM. Consequently, these weak-scattering propagations effectively dilute the time-averaged global pitch-angle scattering rate.

The physical mechanism driving the enhanced perpendicular transport in the multiphase ISM involves two complementary processes operating in distinct spatial regimes. Throughout the volume-filling WNM and UNM, pitch-angle scattering is weak (see Fig.~\ref{fig:Dmu}), yielding a parallel mean free path $\lambda_\parallel$ that exceeds the turbulence injection scale. In this regime, CRs effectively trace the stochastic separation of magnetic field lines, and the perpendicular transport follows the superdiffusion statistics \citep{2014ApJ...784...38L,2022MNRAS.512.2111H}. Because the global $D_{\mu\mu}$ in the multiphase ISM is smaller than in the isothermal case (see Fig.~\ref{fig:Dmu}), particles retain long streaming segments between scattering events, preserving the conditions for field-line-wandering-dominated perpendicular superdiffusion over most of their trajectory.

At the thermal phase-transition boundaries, localized cross-field transport is further enhanced by two mechanisms that depend on the local adiabaticity parameter $\epsilon \equiv |\nabla B|\, r_L / B$. The magnetic field strength varies on spatial scales comparable to or smaller than the particle Larmor radius, as evidenced by the extreme gradients $|\nabla B| \sim 1$--$100~\SI{}{\micro G}$~pc$^{-1}$ (Fig.~\ref{fig:gradT_gradB}). For PeV particles with $r_L \sim 0.5$--$5$~pc, $\epsilon$ reaches values of $\epsilon \sim 0.2$--$170$ at the phase boundaries where significant scattering occurs (Fig.~\ref{fig:T_B_mu}). The scattering events that dominate the cross-field transport---those with $|d\mu/d(t\Omega)| \gtrsim 0.1$---are concentrated at $|dB/d(t\Omega)| \sim 1$--$10~\SI{}{\micro G}$, corresponding to $\epsilon \sim 0.3$--$3$ (using $\epsilon = |dB/d(t\Omega)|/B$ with $B \sim 3~\SI{}{\micro G}$). In this regime, the first adiabatic invariant $\mu_\mathrm{mag} = m v_\perp^2 / 2B$ is violated, and the particle's guiding center undergoes a finite perpendicular displacement during the scattering event itself. For comparison, the isothermal medium also exhibits scattering events extending to $|dB/d(t\Omega)| \sim 10^1~\SI{}{\micro G}$ (Fig.~\ref{fig:B_mu}), but these are statistically rare---the peak scattering occurs at $|dB/d(t\Omega)| \sim 10^{-1}~\SI{}{\micro G}$, corresponding to $\epsilon \sim 0.03$, where the classical mirror diffusion framework \citep{2021ApJ...923...53L} applies. Thermal phase transitions fundamentally reshape the $|\nabla B|$ distribution by systematically concentrating steep magnetic gradients at phase boundaries, elevating the frequency of $\epsilon \gtrsim 1$ events by orders of magnitude and shifting the statistical weight of the scattering toward the non-adiabatic regime, where direct guiding-center cross-field displacements provide a qualitatively more efficient perpendicular transport channel.

The coexistence of these two spatial regimes explains the seemingly paradoxical result that $D_\perp$ is dramatically enhanced while the global $D_{\mu\mu}$ is simultaneously \textit{reduced} relative to the isothermal case. The reduced $D_{\mu\mu}$ reflects particles spending most of their trajectory in the weakly scattering WNM/UNM, where they stream along wandering field lines and accumulate perpendicular displacement via field line separation. The non-adiabatic events at the phase boundaries then inject additional discrete cross-field jumps that further decorrelate the particle from any single magnetic line. In the isothermal medium, by contrast, the scattering distribution is statistically dominated by adiabatic events ($\epsilon \ll 1$), and perpendicular transport relies primarily on the slower process of field line wandering modulated by mirror bounces. Together, the field line separation in the diffuse phases and the frequent non-adiabatic cross-field scattering at the phase boundaries drive $D_\perp$ toward $D_\parallel$, producing the nearly isotropic spatial diffusion seen in Fig.~\ref{fig:DE}. The relative importance of the non-adiabatic channel increases with particle energy, as higher-energy CRs with larger $r_L$ yield larger $\epsilon$ at a given $|\nabla B|$.

\subsection{Cosmic ray diffusion in multi-phase ISM}
\subsubsection{Total spatial and pitch-angle diffusion coefficients}
\begin{figure*}
\centering
\includegraphics[width=0.99\linewidth]{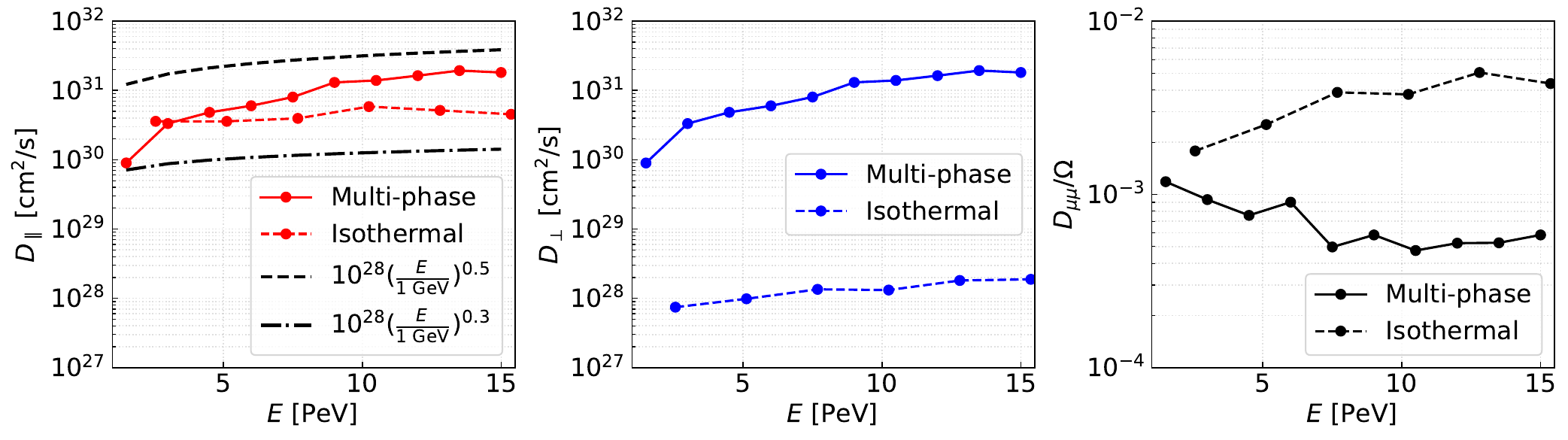}
        \caption{ Energy dependence of the cosmic-ray diffusion coefficients in the simulated interstellar medium. The panels, from left to right, display the parallel spatial diffusion coefficient ($D_\parallel$), the perpendicular spatial diffusion coefficient ($D_\perp$), and the pitch-angle diffusion coefficient normalized by the gyrofrequency ($D_{\mu\mu}/\Omega$) as a function of cosmic-ray energy $E$ in PeV. The initial pitch angle is randomly and isotropically distributed. In all panels, the solid lines represent the results from the realistic multi-phase ISM simulation, while the dashed lines with a circular symbol denote the isothermal (single-phase) control run. In the left panel, the dashed and dash-dotted lines represent the empirical parallel diffusion coefficient derived from CR data fitting.}
    \label{fig:DE}
\end{figure*}

Fig.~\ref{fig:DE} illustrates the energy dependence of the spatial and pitch-angle diffusion coefficients for cosmic rays propagating through the ISM. Assuming an initially isotropic pitch-angle distribution, the parallel spatial diffusion coefficient ($D_\parallel$, left panel) increases steadily by more than an order of magnitude, rising from $\sim 10^{30}\ \rm{cm^2\ s^{-1}}$ at 1.5 PeV to roughly $2 \times 10^{31}\ \rm{cm^2\ s^{-1}}$ at 15 PeV. Similarly, the perpendicular spatial diffusion coefficient ($D_\perp$, middle panel) mirrors this scaling, growing concurrently from $\sim 10^{30}\ \rm{cm^2\ s^{-1}}$ to nearly $2 \times 10^{31}\ \rm{cm^2\ s^{-1}}$. For context, a standard empirical parametrization of the CR parallel diffusion coefficient derived from observational data fitting is typically expressed as:
\begin{equation}
 D_{\rm \parallel,~emp.} \sim (10^{28}\text{--}10^{29}) \left(\frac{E}{\rm 1~GeV}\right)^s \ \rm{cm^2\ s^{-1}},
\end{equation}
with $s \sim 0.3\text{--}0.6$ \citep{2007ARNPS..57..285S}. As shown in Fig.~\ref{fig:DE}, our simulated $D_\parallel$ falls within the range bracketed by $10^{28} (E/{\rm 1~GeV})^{0.3} \ \rm{cm^2\ s^{-1}}$ and $10^{28} (E/{\rm 1~GeV})^{0.5}\ \rm{cm^2\ s^{-1}}$.

This trend is intrinsically linked to the multiphase properties of the ISM. CRs with smaller Larmor radii more frequently sample the highly structured phase-transition boundaries between the WNM/UNM and CNM. Furthermore, the super-Alfv\'enic nature of the CNM dictates that CRs with a Larmor radius smaller than the characteristic size of these cold clumps experience intense, localized scattering, resulting in a slower diffusion rate within the cold phase. Accordingly, the normalized pitch-angle scattering rate ($D_{\mu\mu}/\Omega$, right panel) shows a general decreasing trend, dropping from $\sim 3 \times 10^{-4}$ down to $\sim 10^{-4}$ toward higher energies. As the Larmor radius increases, it becomes more difficult for particles to deeply interact with the sharp, small-scale magnetic field fluctuations at phase boundaries, naturally leading to larger spatial mean free paths.

When comparing these multiphase results with the single-phase isothermal medium, spatial diffusion remains relatively flat and insensitive to particle energy, with $D_\parallel$ hovering around $3-4 \times 10^{30}\ \rm{cm^2\ s^{-1}}$ and $D_\perp$ stagnating near $\sim 10^{28}\ \rm{cm^2\ s^{-1}}$. This weak energy dependence in the isothermal case within this energy range is expected. 
The value suggests a parallel mean free path $\lambda_\parallel$ of $\sim100$~pc, assuming $D_\parallel\approx \frac{1}{3}\lambda_\parallel c$, which means the scattering appears relatively weak within the simulation box. 
The multiphase environment enhances the perpendicular diffusion ($D_\perp$) by roughly two to three orders of magnitude—jumping from $\sim 10^{28}\ \rm{cm^2\ s^{-1}}$ in the isothermal case to $\sim 10^{30}-10^{31}\ \rm{cm^2\ s^{-1}}$ in the multiphase regime. This enhancement occurs even though the local pitch-angle scattering rate in the multiphase ISM ($D_{\mu\mu}/\Omega \sim 5\times 10^{-4} -10^{-3}$) is suppressed by roughly a factor of 2 at 2.5 PeV and 10 at 15 PeV relative to the isothermal medium ($D_{\mu\mu}/\Omega \sim 2 \times 10^{-3} - 5 \times 10^{-3}$). This difference suggests that the enhanced perpendicular diffusion in the multiphase ISM is fundamentally governed by strong scattering at the thermal phase boundaries. As particles encounter significant magnetic fluctuations at the phase-transition boundaries, the resulting strong scattering forces them to deviate from their original field lines and jump to adjacent ones. When coupled with the wandering of magnetic field lines \citep{2014ApJ...784...38L,2013ApJ...779..140X,2022MNRAS.512.2111H,2025ApJ...994..142H}—an intrinsic feature of MHD turbulence—these cross-field jumps efficiently isotropize the spatial transport, driving $D_\perp$ to approach $D_\parallel$.

\begin{figure*}
\centering
\includegraphics[width=0.99\linewidth]{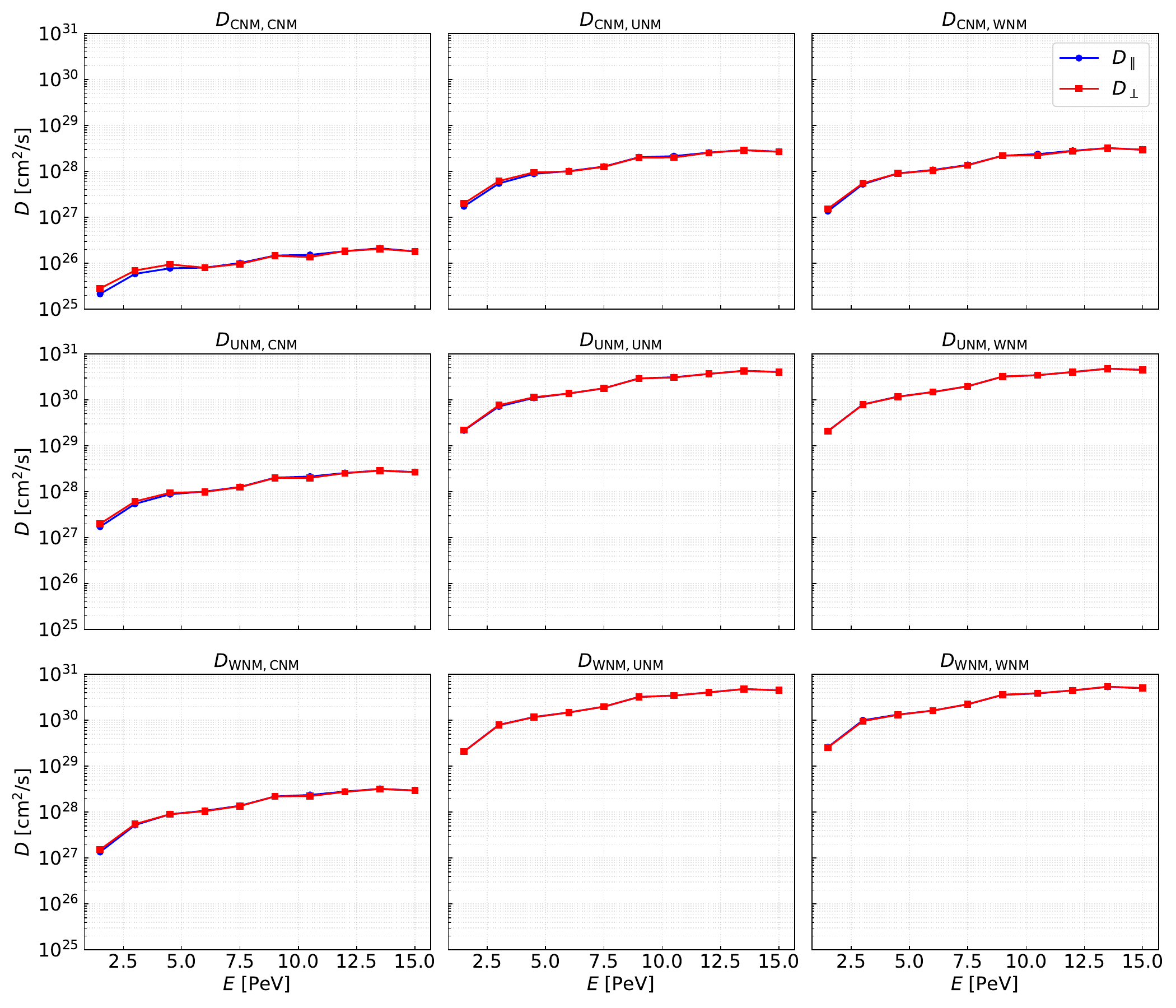}
        \caption{Phase--phase decomposition of the cosmic-ray diffusion coefficients as a function of energy $E$. The $3 \times 3$ matrix displays the components of the parallel ($D_{\alpha\beta}^{\parallel}$, blue lines) and perpendicular ($D_{\alpha\beta}^{\perp}$, red lines) diffusion coefficient matrices for the CNM, UNM, and WNM. As defined in Eqs.~\ref{eq:Dpar} and \ref{eq:Dperp}, the diagonal elements ($D_{\alpha\alpha}$) quantify the intra-phase diffusion contributions, while the off-diagonal elements ($D_{\alpha\beta}$) capture the cross-correlations between displacements accumulated in different phases.}
    \label{fig:Dmatrix}
\end{figure*}

\begin{figure*}
\centering
\includegraphics[width=0.99\linewidth]{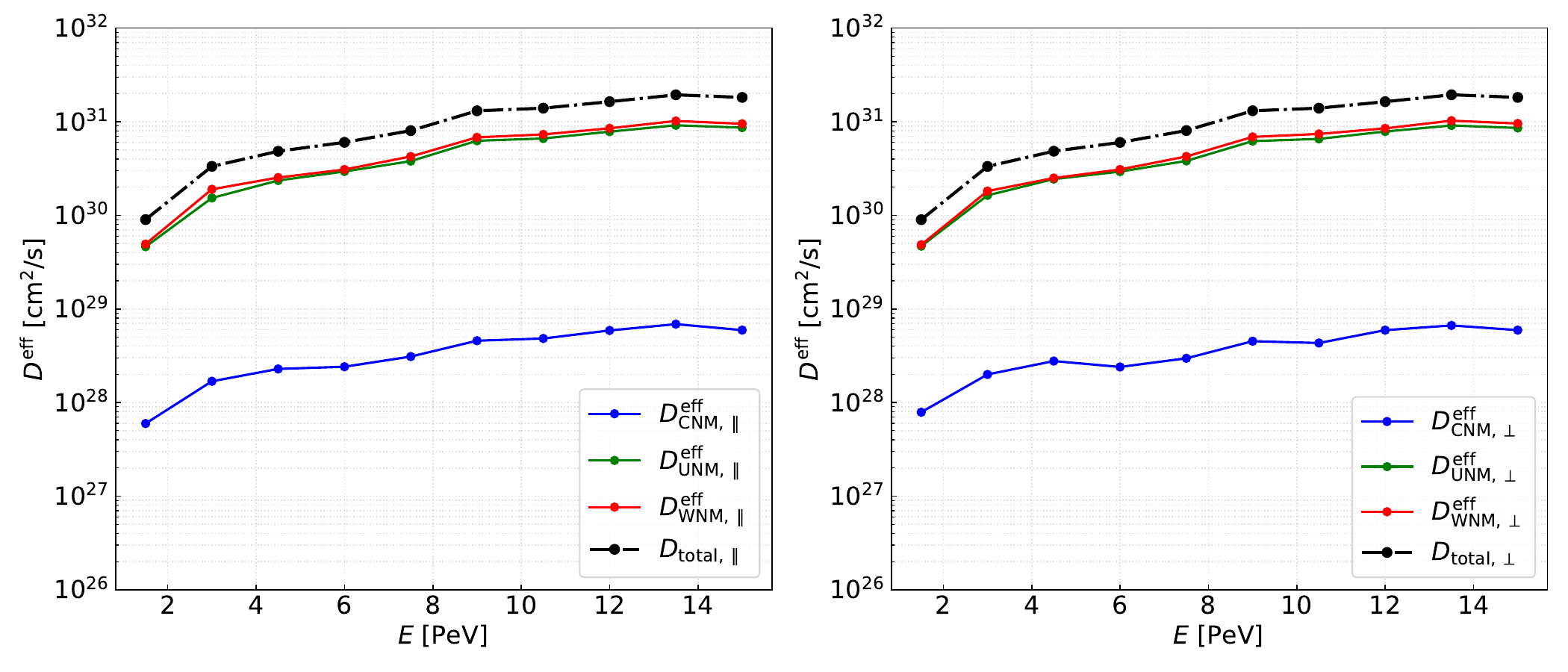}
        \caption{Effective phase-specific diffusion coefficients as a function of cosmic-ray energy $E$. The left and right panels display the parallel ($D^{\text{eff}}_{\parallel}$) and perpendicular ($D^{\text{eff}}_{\perp}$) components, respectively. These effective coefficients are derived from the diagonal elements of the phase--phase diffusion matrix ($D_{\alpha\alpha}$) normalized by the fractional time the particles spend in each respective phase (which tracks the volume-filling factor along the total trajectory). The black dot-dashed line represents the total macroscopic diffusion coefficient. The intrinsic diffusion within the WNM (red) and UNM (green) phases is similar and governs the overall transport rate. In stark contrast, the normalized effective diffusion within the CNM (blue) remains orders of magnitude lower across all energies.}
    \label{fig:Deff}
\end{figure*}
\subsubsection{Phase-decomposed spatial diffusion coefficients}

To quantify the exact contribution of each thermal phase, Fig.~\ref{fig:Dmatrix} presents the decomposed phase-phase diffusion matrix ($D_{\alpha\beta}$). The total diffusion is dominated by the volume-filling WNM and UNM phases, as evidenced by their intra-phase contributions ($D_{\text{WNM, WNM}}$ and $D_{\text{UNM, UNM}}$), which scale from $\sim 10^{29}\text{ cm}^2\text{ s}^{-1}$ at $1.5\text{ PeV}$ up to $\sim 10^{31}\text{ cm}^2\text{ s}^{-1}$ at $15\text{ PeV}$. In contrast, the CNM intra-phase contribution ($D_{\text{CNM, CNM}}$) is bounded between $\sim 10^{25}$ and $10^{26}\text{ cm}^2\text{ s}^{-1}$. Crucially, all off-diagonal cross-terms are strictly positive ($D_{\alpha\beta} > 0$), demonstrating that particle drifts across different thermal phases are positively correlated and cooperatively enhance the global diffusion rate.

To remove the bias introduced by the disparate volume fractions of each phase, Fig.~\ref{fig:Deff} displays the effective phase-specific diffusion coefficients ($D^{\text{eff}}$), normalized by the fractional trajectory time spent in each phase. The results show a small CR diffusion coefficient in the cold phase: the effective diffusion within the CNM (blue lines, $D^{\text{eff}}_{\text{CNM}} \sim 6\times10^{27}-6\times10^{28}\text{ cm}^2\text{ s}^{-1}$) remains two to three orders of magnitude lower than in the WNM and UNM ($D^{\text{eff}}_{\text{WNM}}, D^{\text{eff}}_{\text{UNM}} \sim 5\times10^{29}-1\times10^{31}\text{ cm}^2\text{ s}^{-1}$) across all tracked energy scales. It confirms that PeV CRs experience prolonged, nearly scatter-free streaming in WNM and UNM within 100~pc, while the weakly magnetized CNM intrinsically acts as an intense scattering medium, impeding both the parallel and cross-field mobility of CRs whenever they penetrate these dense structures.

\section{Discussion}
\label{sec:discussion}
\subsection{Comparison with earlier work}
The transport of CRs in turbulent magnetic fields has historically been modeled using Quasi-Linear Theory (QLT) and its non-linear extensions, which predominantly assume a single-phase background plasma \citep{1966ApJ...146..480J,1969ApJ...155..777J,2002PhRvL..89B1102Y,2008ApJ...673..942Y,2021ApJ...923...53L,2023MNRAS.525.4985K,2023JPlPh..89e1701L,2026MNRAS.545f2108E}. These standard models predict that spatial diffusion is governed by the magnetic field line wandering and continuous pitch-angle scattering rate ($D_{\mu\mu}$). However, our results demonstrate a paradigm shift when introducing multiphase thermodynamics. By comparing the multiphase ISM to an isothermal control run, we showed that the global, volume-averaged $D_{\mu\mu}$ remains at similar or even lower levels, thereby preserving a large $D_\parallel$. Yet, the perpendicular diffusion ($D_\perp$) is enhanced by several orders of magnitude. This apparent paradox confirms that in a multiphase ISM, perpendicular spatial transport is not merely a diffusive process driven by continuous background scattering. Rather, it is dominantly driven by the synergy of turbulent magnetic field line wandering and discrete scattering events---both adiabatic mirror reflections and non-adiabatic interactions---localized at thermal phase boundaries.


\subsection{Observation implications}
\subsubsection{Isotropic diffusion}
Recent observations of highly symmetric, extended TeV--PeV gamma-ray halos around Galactic sources by HAWC \citep{2017Sci...358..911A} and LHAASO \citep{2021Natur.594...33C} strongly imply that the diffusion of high-energy CRs in the local ISM is roughly isotropic ($D_\perp \sim D_\parallel$). IceCube \citep{2013ApJ...765...55A, 2016ApJ...826..220A} and Tibet AS$\gamma$ \citep{2006Sci...314..439A} measurements of the large-scale CR anisotropy at TeV--PeV energies, with dipole amplitudes at the $\sim 10^{-3}$ level, provide additional constraints on the diffusion tensor, though the dipole amplitude depends on both $D_\perp / D_\parallel$ and the large-scale CR density gradient.
However, standard CR transport theory in isothermal media generally gives a slower perpendicular diffusion. If applied to local Galactic sources, this highly anisotropic diffusion ($D_\perp \ll D_\parallel$, see Fig.~\ref{fig:DE}) would constrain the CRs to a tube-like geometry, producing highly asymmetric, elongated halos unless the local mean magnetic field is preferentially aligned with the observer's line of sight \citep{2019PhRvL.123v1103L, 2022PhRvD.106l3033D}.

Our findings naturally provide a solution to this problem: (i) the stochastic wandering of magnetic field lines in the turbulent medium causes CRs streaming along those lines to undergo perpendicular superdiffusion relative to the mean magnetic field; (ii) the multiphase nature of the ISM intrinsically induces significant magnetic field fluctuations at phase boundaries, enhancing pitch angle scattering and cross-field jumping. Combining the two effects enhances $D_\perp$ to be comparable with $D_\parallel$, providing the physical foundation required to support the observed isotropic morphologies of Galactic PeVatrons. We expect the isotropic diffusion to also be energy dependent, as low-energy CRs with a smaller Larmor radius experience a smoother transition at the phase boundaries. 

\subsubsection{Slow diffusion in CNM}
We show that the PeV CRs' diffusion in CNM is slower than in WNM and UNM. This slow diffusion naturally comes from the super-Alfv\'enic properties of CNM, which means the magnetic field fluctuations are stronger than those in sub-Alfv\'enic/tran-Alfv\'enic WNM and UNM. CNM intrinsically acts as a confining environment for PeV particles. Since $\gtrsim 1 \text{ PeV}$ gamma-ray emissions are primarily produced via hadronic (proton-proton) interactions, the gamma-ray luminosity is proportional to the product of the CR number density and the local background gas density. This confinement effect predicts that the hadronic gamma-ray morphology around PeVatrons should be modulated by the local phase structure of the ISM: rather than a smooth, symmetric halo, one expects enhanced emission from dense CNM clumps that efficiently trap PeV CRs, potentially contributing to the clumpy substructure observed in some LHAASO sources.

It is important to note that the diffusion coefficients inferred from HAWC and LHAASO observations of TeV–PeV halos \citep{2017Sci...358..911A,2021Natur.594...33C} are measured in the immediate vicinity of individual sources, where the local ISM conditions---including possible self-confinement by CR-driven instabilities, enhanced turbulence from the progenitor system, or a typically dense ambient gas---may strongly suppress transport relative to the Galactic average. Our simulation, by contrast, characterizes the \textit{global} diffusion properties of PeV CRs propagating through a statistically representative, turbulence-driven multiphase ISM, yielding $D_\parallel \sim D_\perp \sim 10^{30}~{\rm cm^2~s^{-1}}$ at 1.5~PeV. The three-order-of-magnitude difference, therefore, does not represent a contradiction; rather, it reflects the distinction between the source-vicinity environment---where localized confinement mechanisms operate---and the large-scale ISM through which CRs propagate after escaping the immediate source region. Future work incorporating a realistic CR source embedded within the multiphase ISM will be needed to bridge these two regimes and directly model the formation of extended gamma-ray halos.

\subsection{Limitations}
Several limitations of the present work should be noted. First, our simulation employs purely solenoidal turbulent driving, whereas realistic ISM energy injection by supernovae includes a significant compressive component that could produce stronger shocks, sharper phase boundaries, and more intense magnetic fluctuations at phase interfaces. 

Second, we adopt a spatially uniform heating rate $\Gamma$ and a simplified atomic cooling function $\Lambda$. While this parameterization captures the essential bistable thermal equilibrium that generates the WNM--UNM--CNM phase structure, it omits processes such as photoelectric heating variations with local radiation field intensity, molecular cooling at high densities, and metallicity dependence---all of which could modify the phase fractions, the sharpness of phase boundaries, and consequently the magnetic gradient statistics. 

Third, our ideal MHD framework does not resolve ion-neutral decoupling in the partially ionized CNM and UNM. Ion-neutral collisions damp MHD fluctuations below the decoupling scale, which would reduce the small-scale magnetic fluctuations available for resonant pitch-angle scattering within the CNM interior \citep{1971ApL.....8..189K,2016ApJ...826..166X,2021ApJ...914....3P,2022FrP....10.2799L,2024MNRAS.527.3945H,2025ApJ...992...10K}. However, the dominant scattering mechanism identified in this work---non-adiabatic interactions at the large-scale magnetic gradients associated with phase boundaries---operates on scales well above the ion-neutral decoupling scale and is therefore expected to be less affected. A quantitative assessment of this effect requires two-fluid (ion--neutral) MHD simulations, which will be addressed in forthcoming work.

\section{Conclusion} 
\label{sec:conclusion}
In this paper, we investigated the propagation of $\sim$\,PeV cosmic rays in a highly turbulent, multiphase interstellar medium. We performed high-resolution ($2048^3$) three-dimensional ideal MHD simulations that self-consistently generate the WNM, UNM, and CNM through the interplay of MHD turbulence and atomic heating/cooling. Using a test-particle tracking approach across a range of particle energies (1.5--15\,PeV), we examined the mechanisms of pitch-angle scattering and spatial diffusion within and between the distinct thermal phases. Our principal conclusions are as follows:

\begin{enumerate}

\item Nonlinear thermal phase transitions in the multiphase ISM generate steep, co-spatial gradients in both gas temperature ($|\nabla T| \sim 10^3$--$10^4$\,K\,pc$^{-1}$) and magnetic field strength ($|\nabla B| \sim 1$--$100\,\SI{}{\micro G}$\,pc$^{-1}$). These localized magnetic fluctuations produce efficient scattering sites for PeV CRs that are absent in isothermal models of the ISM.
\item A strong positive correlation exists between the rates of change of the pitch-angle cosine $|d\mu/d(t\Omega)|$, the temperature $|dT/d(t\Omega)|$, and the magnetic field strength $|dB/d(t\Omega)|$ along particle trajectories. This demonstrates that the most intense scattering events---characterized by large, abrupt pitch-angle changes---are directly driven by particles traversing the extreme magnetic gradients at thermal phase-transition boundaries.
\item  The global pitch-angle diffusion coefficient in the multiphase medium is approximately a factor of two smaller than in an equivalent isothermal medium, reflecting the fact that particles spend the majority of their trajectory in the weakly scattering, volume-filling WNM and UNM. Despite this reduced scattering rate, the perpendicular spatial diffusion coefficient ($D_\perp$) is amplified by two to three orders of magnitude relative to the isothermal case, becoming comparable to the parallel coefficient ($D_\parallel \sim D_\perp \sim 10^{30}$,cm$^2$,s$^{-1}$ at 1.5,PeV). This enhancement arises from three cooperating mechanisms: (i) field line separation and superdiffusion in the weakly scattering WNM/UNM, where the parallel mean free path ($\lambda_\parallel \sim 100$,pc) exceeds the turbulence injection scale; (ii) adiabatic mirror diffusion at phase boundaries where the adiabaticity parameter $\epsilon = |\nabla B|,r_L/B \lesssim 1$; and (iii) non-adiabatic cross-field scattering at the sharpest boundaries where $\epsilon \gtrsim 1$ and the first adiabatic invariant is violated. Together, these processes drive the nearly isotropic spatial diffusion.
\item Using a phase--phase diffusion matrix decomposition, we show that the volume-filling WNM and UNM primarily govern PeV CR transport, with intra-phase contributions scaling from $\sim 10^{29}$\,cm$^2$\,s$^{-1}$ at 1.5\,PeV to $\sim 10^{31}$\,cm$^2$\,s$^{-1}$ at 15\,PeV. The cross-correlation terms ($D_{\alpha\beta}$) between different thermal phases are universally positive, indicating that particle drifts across phase interfaces cooperatively enhance the global spatial transport.
\item After normalizing by the fractional trajectory time spent in each phase, the effective diffusion coefficient within the CNM ($D_\mathrm{CNM}^\mathrm{eff} \sim 6 \times 10^{27}$--$6 \times 10^{28}$\,cm$^2$\,s$^{-1}$) is two to three orders of magnitude smaller than in the WNM and UNM ($D_\mathrm{WNM}^\mathrm{eff},\, D_\mathrm{UNM}^\mathrm{eff} \sim 5 \times 10^{29}$--$10^{31}$\,cm$^2$\,s$^{-1}$) across the energy range 1.5--15\,PeV. The highly compressive and super-Alfv\'enic cold structures act as efficient trapping environments that substantially suppress both parallel and perpendicular diffusion of PeV CRs.
\item Our results yield two observational implications. First, the nearly isotropic diffusion ($D_\perp \sim D_\parallel$) obtained in the global multiphase ISM provides a necessary physical condition for explaining the symmetric, extended TeV--PeV halos observed around Galactic sources. While the diffusion coefficients measured in the immediate source vicinity by HAWC and LHAASO ($D \sim 10^{26}$--$10^{27}$\,cm$^2$\,s$^{-1}$) reflect local confinement mechanisms not captured by our global ISM simulation, the isotropy of the bulk ISM diffusion tensor constrains the large-scale halo morphology once CRs escape the source region. Second, the strong CR confinement within the CNM predicts that the hadronic gamma-ray brightness should be modulated by the local ISM phase structure, with enhanced emission from dense CNM clumps where the CR energy density is amplified. This clumpy substructure provides a testable signature for future high-resolution LHAASO and CTA observations.
\end{enumerate}

These results highlight the importance of incorporating realistic multiphase thermodynamics into models of Galactic CR transport, and provide a theoretical framework for interpreting the spatial morphology and substructure of the diffuse PeV gamma-ray emission observed from Galactic PeVatrons.
\vspace{6pt}

\dataavailability{The data underlying this article will be shared on reasonable request to the corresponding author.}

\acknowledgments{Y.H. acknowledges the support for this work provided by NASA through the NASA Hubble Fellowship grant No. HST-HF2-51557.001 awarded by the Space Telescope Science Institute, which is operated by the Association of Universities for Research in Astronomy, Incorporated, under NASA contract NAS5-26555. This work used SDSC Expanse CPU and NCSA Delta CPU through allocations PHY230032, PHY230033, PHY230091, PHY230105,  PHY230178, PHY240188, and PHY240183, from the Advanced Cyberinfrastructure Coordination Ecosystem: Services \& Support (ACCESS) program, which is supported by National Science Foundation grants \#2138259, \#2138286, \#2138307, \#2137603, and \#2138296. }

\conflictsofinterest{The authors declare no conflicts of interest.} 



\abbreviations{Abbreviations}{
The following abbreviations are used in this manuscript:
\\

\noindent 
\begin{tabular}{@{}ll}
MHD & Magnetohydrodynamic\\
ISM & Interstellar Medium\\
CR & Cosmic Ray\\
WNM & Warm Neutral Medium\\
UNM & Unstable Neutral Medium\\
CNM & Cold Neutral Medium\\
LHAASO & Large High Altitude Air Shower Observatory\\
MSD & Mean-Squared Displacement
\end{tabular}
}




\begin{adjustwidth}{-\extralength}{0cm}

\reftitle{References}


\bibliography{template}

\begin{thebibliography}{999}

\bibitem[{Strong} et~al.(2007){Strong}, {Moskalenko}, and {Ptuskin}]{2007ARNPS..57..285S}
{Strong}, A.W.; {Moskalenko}, I.V.; {Ptuskin}, V.S.
\newblock {Cosmic-Ray Propagation and Interactions in the Galaxy}.
\newblock {\em Annual Review of Nuclear and Particle Science} {\bf 2007}, {\em 57},~285--327,  \href{http://arxiv.org/abs/astro-ph/0701517}{{\normalfont [arXiv:astro-ph/astro-ph/0701517]}}.
\newblock {\url{https://doi.org/10.1146/annurev.nucl.57.090506.123011}}.

\bibitem[{Everett} et~al.(2008){Everett}, {Zweibel}, {Benjamin}, {McCammon}, {Rocks}, and {Gallagher}]{2008ApJ...674..258E}
{Everett}, J.E.; {Zweibel}, E.G.; {Benjamin}, R.A.; {McCammon}, D.; {Rocks}, L.; {Gallagher}, III, J.S.
\newblock {The Milky Way's Kiloparsec-Scale Wind: A Hybrid Cosmic-Ray and Thermally Driven Outflow}.
\newblock {\em \apj} {\bf 2008}, {\em 674},~258--270,  \href{http://arxiv.org/abs/0710.3712}{{\normalfont [arXiv:astro-ph/0710.3712]}}.
\newblock {\url{https://doi.org/10.1086/524766}}.

\bibitem[{Jubelgas} et~al.(2008){Jubelgas}, {Springel}, {En{\ss}lin}, and {Pfrommer}]{2008A&A...481...33J}
{Jubelgas}, M.; {Springel}, V.; {En{\ss}lin}, T.; {Pfrommer}, C.
\newblock {Cosmic ray feedback in hydrodynamical simulations of galaxy formation}.
\newblock {\em \aap} {\bf 2008}, {\em 481},~33--63,  \href{http://arxiv.org/abs/astro-ph/0603485}{{\normalfont [arXiv:astro-ph/astro-ph/0603485]}}.
\newblock {\url{https://doi.org/10.1051/0004-6361:20065295}}.

\bibitem[Van~Rossum and Drake(2009)]{10.5555/1593511}
Van~Rossum, G.; Drake, F.L.
\newblock {\em Python 3 Reference Manual}; CreateSpace: Scotts Valley, CA,  2009.

\bibitem[{VERITAS Collaboration} et~al.(2009){VERITAS Collaboration}, {Acciari}, {Aliu}, {Arlen}, {Aune}, {Bautista}, {Beilicke}, {Benbow}, {Boltuch}, {Bradbury}, {Buckley}, {Bugaev}, {Byrum}, {Cannon}, {Celik}, {Cesarini}, {Chow}, {Ciupik}, {Cogan}, {Colin}, {Cui}, {Dickherber}, {Duke}, {Fegan}, {Finley}, {Finnegan}, {Fortin}, {Fortson}, {Furniss}, {Galante}, {Gall}, {Gibbs}, {Gillanders}, {Godambe}, {Grube}, {Guenette}, {Gyuk}, {Hanna}, {Holder}, {Horan}, {Hui}, {Humensky}, {Imran}, {Kaaret}, {Karlsson}, {Kertzman}, {Kieda}, {Kildea}, {Konopelko}, {Krawczynski}, {Krennrich}, {Lang}, {Lebohec}, {Maier}, {McArthur}, {McCann}, {McCutcheon}, {Millis}, {Moriarty}, {Mukherjee}, {Nagai}, {Ong}, {Otte}, {Pandel}, {Perkins}, {Pizlo}, {Pohl}, {Quinn}, {Ragan}, {Reyes}, {Reynolds}, {Roache}, {Rose}, {Schroedter}, {Sembroski}, {Smith}, {Steele}, {Swordy}, {Theiling}, {Thibadeau}, {Varlotta}, {Vassiliev}, {Vincent}, {Wagner}, {Wakely}, {Ward}, {Weekes}, {Weinstein}, {Weisgarber}, {Williams}, {Wissel}, {Wood}, and
  {Zitzer}]{2009Natur.462..770V}
{VERITAS Collaboration}.; {Acciari}, V.A.; {Aliu}, E.; {Arlen}, T.; {Aune}, T.; {Bautista}, M.; {Beilicke}, M.; {Benbow}, W.; {Boltuch}, D.; {Bradbury}, S.M.;  et~al.
\newblock {A connection between star formation activity and cosmic rays in the starburst galaxy M82}.
\newblock {\em \nat} {\bf 2009}, {\em 462},~770--772,  \href{http://arxiv.org/abs/0911.0873}{{\normalfont [arXiv:astro-ph.CO/0911.0873]}}.
\newblock {\url{https://doi.org/10.1038/nature08557}}.

\bibitem[{Girichidis} et~al.(2016){Girichidis}, {Naab}, {Walch}, {Hanasz}, {Mac Low}, {Ostriker}, {Gatto}, {Peters}, {W{\"u}nsch}, {Glover}, {Klessen}, {Clark}, and {Baczynski}]{2016ApJ...816L..19G}
{Girichidis}, P.; {Naab}, T.; {Walch}, S.; {Hanasz}, M.; {Mac Low}, M.M.; {Ostriker}, J.P.; {Gatto}, A.; {Peters}, T.; {W{\"u}nsch}, R.; {Glover}, S.C.O.;  et~al.
\newblock {Launching Cosmic-Ray-driven Outflows from the Magnetized Interstellar Medium}.
\newblock {\em \apjl} {\bf 2016}, {\em 816},~L19,  \href{http://arxiv.org/abs/1509.07247}{{\normalfont [arXiv:astro-ph.GA/1509.07247]}}.
\newblock {\url{https://doi.org/10.3847/2041-8205/816/2/L19}}.

\bibitem[{Ruszkowski} et~al.(2017){Ruszkowski}, {Yang}, and {Zweibel}]{2017ApJ...834..208R}
{Ruszkowski}, M.; {Yang}, H.Y.K.; {Zweibel}, E.
\newblock {Global Simulations of Galactic Winds Including Cosmic-ray Streaming}.
\newblock {\em \apj} {\bf 2017}, {\em 834},~208,  \href{http://arxiv.org/abs/1602.04856}{{\normalfont [arXiv:astro-ph.GA/1602.04856]}}.
\newblock {\url{https://doi.org/10.3847/1538-4357/834/2/208}}.

\bibitem[{Hopkins} et~al.(2021){Hopkins}, {Chan}, {Ji}, {Hummels}, {Kere{\v{s}}}, {Quataert}, and {Faucher-Gigu{\`e}re}]{2021MNRAS.501.3640H}
{Hopkins}, P.F.; {Chan}, T.K.; {Ji}, S.; {Hummels}, C.B.; {Kere{\v{s}}}, D.; {Quataert}, E.; {Faucher-Gigu{\`e}re}, C.A.
\newblock {Cosmic ray driven outflows to Mpc scales from L$_{*}$ galaxies}.
\newblock {\em \mnras} {\bf 2021}, {\em 501},~3640--3662,  \href{http://arxiv.org/abs/2002.02462}{{\normalfont [arXiv:astro-ph.GA/2002.02462]}}.
\newblock {\url{https://doi.org/10.1093/mnras/staa3690}}.

\bibitem[{Liu} et~al.(2022){Liu}, {Hu}, and {Lazarian}]{2022MNRAS.510.4952L}
{Liu}, M.; {Hu}, Y.; {Lazarian}, A.
\newblock {Velocity gradients: magnetic field tomography towards the supernova remnant W44}.
\newblock {\em \mnras} {\bf 2022}, {\em 510},~4952--4961,  \href{http://arxiv.org/abs/2109.13670}{{\normalfont [arXiv:astro-ph.GA/2109.13670]}}.
\newblock {\url{https://doi.org/10.1093/mnras/stab3783}}.

\bibitem[{Krumholz} et~al.(2023){Krumholz}, {Crocker}, and {Offner}]{2023MNRAS.520.5126K}
{Krumholz}, M.R.; {Crocker}, R.M.; {Offner}, S.S.R.
\newblock {The cosmic ray ionization and {\ensuremath{\gamma}}-ray budgets of star-forming galaxies}.
\newblock {\em \mnras} {\bf 2023}, {\em 520},~5126--5143,  \href{http://arxiv.org/abs/2211.03488}{{\normalfont [arXiv:astro-ph.GA/2211.03488]}}.
\newblock {\url{https://doi.org/10.1093/mnras/stad459}}.

\bibitem[{Hopkins} et~al.(2025){Hopkins}, {Quataert}, {Ponnada}, and {Silich}]{2025arXiv250118696H}
{Hopkins}, P.F.; {Quataert}, E.; {Ponnada}, S.B.; {Silich}, E.
\newblock {Cosmic Rays Masquerading as Hot CGM Gas: An Inverse-Compton Origin for Diffuse X-ray Emission in the Circumgalactic Medium}.
\newblock {\em arXiv e-prints} {\bf 2025}, p. arXiv:2501.18696,  \href{http://arxiv.org/abs/2501.18696}{{\normalfont [arXiv:astro-ph.HE/2501.18696]}}.
\newblock {\url{https://doi.org/10.48550/arXiv.2501.18696}}.

\bibitem[{Gaisser}(1990)]{1990cup..book.....G}
{Gaisser}, T.K.
\newblock {\em {Cosmic rays and particle physics.}};  1990.

\bibitem[{Gaisser} et~al.(2013){Gaisser}, {Stanev}, and {Tilav}]{2013FrPhy...8..748G}
{Gaisser}, T.K.; {Stanev}, T.; {Tilav}, S.
\newblock {Cosmic ray energy spectrum from measurements of air showers}.
\newblock {\em Frontiers of Physics} {\bf 2013}, {\em 8},~748--758,  \href{http://arxiv.org/abs/1303.3565}{{\normalfont [arXiv:astro-ph.HE/1303.3565]}}.
\newblock {\url{https://doi.org/10.1007/s11467-013-0319-7}}.

\bibitem[{Abeysekara} et~al.(2017){Abeysekara}, {Albert}, {Alfaro}, {Alvarez}, {{\'A}lvarez}, {Arceo}, {Arteaga-Vel{\'a}zquez}, {Avila Rojas}, {Ayala Solares}, {Barber}, {Bautista-Elivar}, {Becerril}, {Belmont-Moreno}, {BenZvi}, {Berley}, {Bernal}, {Braun}, {Brisbois}, {Caballero-Mora}, {Capistr{\'a}n}, {Carrami{\~n}ana}, {Casanova}, {Castillo}, {Cotti}, {Cotzomi}, {Couti{\~n}o de Le{\'o}n}, {De Le{\'o}n}, {De la Fuente}, {Dingus}, {DuVernois}, {D{\'\i}az-V{\'e}lez}, {Ellsworth}, {Engel}, {Enr{\'\i}quez-Rivera}, {Fiorino}, {Fraija}, {Garc{\'\i}a-Gonz{\'a}lez}, {Garfias}, {Gerhardt}, {Gonz{\'a}lez Mu{\~n}oz}, {Gonz{\'a}lez}, {Goodman}, {Hampel-Arias}, {Harding}, {Hern{\'a}ndez}, {Hern{\'a}ndez-Almada}, {Hinton}, {Hona}, {Hui}, {H{\"u}ntemeyer}, {Iriarte}, {Jardin-Blicq}, {Joshi}, {Kaufmann}, {Kieda}, {Lara}, {Lauer}, {Lee}, {Lennarz}, {Vargas}, {Linnemann}, {Longinotti}, {Luis Raya}, {Luna-Garc{\'\i}a}, {L{\'o}pez-Coto}, {Malone}, {Marinelli}, {Martinez}, {Martinez-Castellanos}, {Mart{\'\i}nez-Castro},
  {Mart{\'\i}nez-Huerta}, {Matthews}, {Miranda-Romagnoli}, {Moreno}, {Mostaf{\'a}}, {Nellen}, {Newbold}, {Nisa}, {Noriega-Papaqui}, {Pelayo}, {Pretz}, {P{\'e}rez-P{\'e}rez}, {Ren}, {Rho}, {Rivi{\`e}re}, {Rosa-Gonz{\'a}lez}, {Rosenberg}, {Ruiz-Velasco}, {Salazar}, {Salesa Greus}, {Sandoval}, {Schneider}, {Schoorlemmer}, {Sinnis}, {Smith}, {Springer}, {Surajbali}, {Taboada}, {Tibolla}, {Tollefson}, {Torres}, {Ukwatta}, {Vianello}, {Weisgarber}, {Westerhoff}, {Wisher}, {Wood}, {Yapici}, {Yodh}, {Younk}, {Zepeda}, {Zhou}, {Guo}, {Hahn}, {Li}, and {Zhang}]{2017Sci...358..911A}
{Abeysekara}, A.U.; {Albert}, A.; {Alfaro}, R.; {Alvarez}, C.; {{\'A}lvarez}, J.D.; {Arceo}, R.; {Arteaga-Vel{\'a}zquez}, J.C.; {Avila Rojas}, D.; {Ayala Solares}, H.A.; {Barber}, A.S.;  et~al.
\newblock {Extended gamma-ray sources around pulsars constrain the origin of the positron flux at Earth}.
\newblock {\em Science} {\bf 2017}, {\em 358},~911--914,  \href{http://arxiv.org/abs/1711.06223}{{\normalfont [arXiv:astro-ph.HE/1711.06223]}}.
\newblock {\url{https://doi.org/10.1126/science.aan4880}}.

\bibitem[{Cao} et~al.(2024){Cao}, {Aharonian}, {Axikegu}, {Bai}, {Bao}, {Bastieri}, {Bi}, {Bi}, {Bian}, {Bukevich}, {Cao}, {Cao}, {Cao}, {Chang}, {Chang}, {Chen}, {Chen}, {Chen}, {Chen}, {Chen}, {Chen}, {Chen}, {Chen}, {Chen}, {Chen}, {Chen}, {Chen}, {Chen}, {Chen}, {Cheng}, {Cheng}, {Cui}, {Cui}, {Cui}, {Cui}, {Dai}, {Dai}, {Dai}, {Danzengluobu}, {Dong}, {Duan}, {Fan}, {Fan}, {Fang}, {Fang}, {Fang}, {Feng}, {Feng}, {Feng}, {Feng}, {Feng}, {Feng}, {Feng}, {Gabici}, {Gao}, {Gao}, {Gao}, {Gao}, {Gao}, {Ge}, {Geng}, {Giacinti}, {Gong}, {Gou}, {Gu}, {Guo}, {Guo}, {Guo}, {Guo}, {Han}, {Hasan}, {He}, {He}, {He}, {He}, {Hor}, {Hou}, {Hou}, {Hou}, {Hu}, {Hu}, {Hu}, {Huang}, {Huang}, {Huang}, {Huang}, {Huang}, {Huang}, {Ji}, {Jia}, {Jia}, {Jiang}, {Jiang}, {Jiang}, {Jin}, {Kang}, {Karpikov}, {Kuleshov}, {Kurinov}, {Li}, {Li}, {Li}, {Li}, {Li}, {Li}, {Li}, {Li}, {Li}, {Li}, {Li}, {Li}, {Li}, {Li}, {Li}, {Li}, {Li}, {Li}, {Li}, {Liang}, {Liang}, {Lin}, {Liu}, {Liu}, {Liu}, {Liu}, {Liu}, {Liu}, {Liu}, {Liu}, {Liu},
  {Liu}, {Liu}, {Liu}, {Liu}, {Liu}, {Luo}, {Luo}, {Lv}, {Ma}, {Ma}, {Ma}, {Mao}, {Min}, {Mitthumsiri}, {Mu}, {Nan}, {Neronov}, {Ou}, {Pattarakijwanich}, {Pei}, {Qi}, {Qi}, {Qiao}, {Qin}, {Raza}, {Ruffolo}, {S{\'a}iz}, {Saeed}, {Semikoz}, {Shao}, {Shchegolev}, {Sheng}, {Shu}, {Song}, {Stenkin}, {Stepanov}, {Su}, {Sun}, {Sun}, {Sun}, {Sun}, {Takata}, {Tam}, {Tang}, {Tang}, {Tang}, {Tian}, {Wang}, {Wang}, {Wang}, {Wang}, {Wang}, {Wang}, {Wang}, {Wang}, {Wang}, {Wang}, {Wang}, {Wang}, {Wang}, {Wang}, {Wang}, {Wang}, {Wang}, {Wang}, {Wang}, {Wang}, {Wang}, {Wang}, and {Wei}]{2024PhRvL.132m1002C}
{Cao}, Z.; {Aharonian}, F.; {Axikegu}.; {Bai}, Y.X.; {Bao}, Y.W.; {Bastieri}, D.; {Bi}, X.J.; {Bi}, Y.J.; {Bian}, W.; {Bukevich}, A.V.;  et~al.
\newblock {Measurements of All-Particle Energy Spectrum and Mean Logarithmic Mass of Cosmic Rays from 0.3 to 30 PeV with LHAASO-KM2A}.
\newblock {\em \prl} {\bf 2024}, {\em 132},~131002,  \href{http://arxiv.org/abs/2403.10010}{{\normalfont [arXiv:astro-ph.HE/2403.10010]}}.
\newblock {\url{https://doi.org/10.1103/PhysRevLett.132.131002}}.

\bibitem[{Blasi}(2013)]{2013A&ARv..21...70B}
{Blasi}, P.
\newblock {The origin of galactic cosmic rays}.
\newblock {\em \aapr} {\bf 2013}, {\em 21},~70,  \href{http://arxiv.org/abs/1311.7346}{{\normalfont [arXiv:astro-ph.HE/1311.7346]}}.
\newblock {\url{https://doi.org/10.1007/s00159-013-0070-7}}.

\bibitem[{Gabici} et~al.(2019){Gabici}, {Evoli}, {Gaggero}, {Lipari}, {Mertsch}, {Orlando}, {Strong}, and {Vittino}]{2019IJMPD..2830022G}
{Gabici}, S.; {Evoli}, C.; {Gaggero}, D.; {Lipari}, P.; {Mertsch}, P.; {Orlando}, E.; {Strong}, A.; {Vittino}, A.
\newblock {The origin of Galactic cosmic rays: Challenges to the standard paradigm}.
\newblock {\em International Journal of Modern Physics D} {\bf 2019}, {\em 28},~1930022--339,  \href{http://arxiv.org/abs/1903.11584}{{\normalfont [arXiv:astro-ph.HE/1903.11584]}}.
\newblock {\url{https://doi.org/10.1142/S0218271819300222}}.

\bibitem[{Marcowith}(2025)]{2025FrASS..1111076M}
{Marcowith}, A.
\newblock {Cosmic rays escape from their sources}.
\newblock {\em Frontiers in Astronomy and Space Sciences} {\bf 2025}, {\em 11},~1411076.
\newblock {\url{https://doi.org/10.3389/fspas.2024.1411076}}.

\bibitem[{Di Mauro} et~al.(2019){Di Mauro}, {Manconi}, and {Donato}]{2019PhRvD.100l3015D}
{Di Mauro}, M.; {Manconi}, S.; {Donato}, F.
\newblock {Detection of a {\ensuremath{\gamma}} -ray halo around Geminga with the Fermi-LAT data and implications for the positron flux}.
\newblock {\em \prd} {\bf 2019}, {\em 100},~123015,  \href{http://arxiv.org/abs/1903.05647}{{\normalfont [arXiv:astro-ph.HE/1903.05647]}}.
\newblock {\url{https://doi.org/10.1103/PhysRevD.100.123015}}.

\bibitem[{Cao} et~al.(2021){Cao}, {Aharonian}, {An}, {Axikegu}, {Bai}, {Bao}, {Bastieri}, {Bi}, {Bi}, {Cai}, {Cai}, {Cao}, {Chang}, {Chang}, {Chang}, {Chen}, {Chen}, {Chen}, {Chen}, {Chen}, {Chen}, {Chen}, {Chen}, {Chen}, {Chen}, {Chen}, {Chen}, {Chen}, {Cheng}, {Cheng}, {Cui}, {Cui}, {Cui}, {Dai}, {Dai}, {Dai}, {Danzengluobu}, {della Volpe}, {D'Ettorre Piazzoli}, {Dong}, {Fan}, {Fan}, {Fan}, {Fang}, {Fang}, {Feng}, {Feng}, {Feng}, {Feng}, {Gao}, {Gao}, {Gao}, {Gao}, {Ge}, {Geng}, {Gong}, {Gou}, {Gu}, {Guo}, {Guo}, {Guo}, {Guo}, {Han}, {He}, {He}, {He}, {He}, {He}, {He}, {Heller}, {Hor}, {Hou}, {Hou}, {Hu}, {Hu}, {Hu}, {Hu}, {Huang}, {Huang}, {Huang}, {Huang}, {Huang}, {Ji}, {Ji}, {Jia}, {Jiang}, {Jiang}, {Jin}, {Kuleshov}, {Levochkin}, {Li}, {Li}, {Li}, {Li}, {Li}, {Li}, {Li}, {Li}, {Li}, {Li}, {Li}, {Li}, {Li}, {Li}, {Li}, {Li}, {Li}, {Liang}, {Liang}, {Lin}, {Liu}, {Liu}, {Liu}, {Liu}, {Liu}, {Liu}, {Liu}, {Liu}, {Liu}, {Liu}, {Liu}, {Liu}, {Liu}, {Liu}, {Liu}, {Long}, {Lu}, {Lv}, {Ma}, {Ma}, {Ma}, {Mao},
  {Masood}, {Mitthumsiri}, {Montaruli}, {Nan}, {Pang}, {Pattarakijwanich}, {Pei}, {Qi}, {Ruffolo}, {Rulev}, {S{\'a}iz}, {Shao}, {Shchegolev}, {Sheng}, {Shi}, {Song}, {Stenkin}, {Stepanov}, {Sun}, {Sun}, {Sun}, {Tam}, {Tang}, {Tian}, {Wang}, {Wang}, {Wang}, {Wang}, {Wang}, {Wang}, {Wang}, {Wang}, {Wang}, {Wang}, {Wang}, {Wang}, {Wang}, {Wang}, {Wang}, {Wang}, {Wang}, {Wang}, {Wang}, {Wang}, {Wang}, {Wei}, {Wei}, {Wei}, {Wen}, {Wu}, {Wu}, {Wu}, {Wu}, {Wu}, {Xi}, {Xia}, {Xia}, {Xiang}, {Xiao}, {Xiao}, {Xin}, {Xin}, {Xing}, {Xu}, {Xu}, {Xue}, {Yan}, and {Yang}]{2021Natur.594...33C}
{Cao}, Z.; {Aharonian}, F.A.; {An}, Q.; {Axikegu}, L.~X., B.; {Bai}, Y.X.; {Bao}, Y.W.; {Bastieri}, D.; {Bi}, X.J.; {Bi}, Y.J.; {Cai}, H.;  et~al.
\newblock {Ultrahigh-energy photons up to 1.4 petaelectronvolts from 12 {\ensuremath{\gamma}}-ray Galactic sources}.
\newblock {\em \nat} {\bf 2021}, {\em 594},~33--36.
\newblock {\url{https://doi.org/10.1038/s41586-021-03498-z}}.

\bibitem[{Lhaaso Collaboration} et~al.(2021){Lhaaso Collaboration}, {Cao}, {Aharonian}, {An}, {Axikegu}, {Bai}, {Bai}, {Bao}, {Bastieri}, {Bi}, {Bi}, {Cai}, {Cai}, {Cao}, {Chang}, {Chang}, {Chen}, {Chen}, {Chen}, {Chen}, {Chen}, {Chen}, {Chen}, {Chen}, {Chen}, {Chen}, {Chen}, {Chen}, {Chen}, {Chen}, {Cheng}, {Cheng}, {Cui}, {Cui}, {Cui}, {D'Ettorre Piazzoli}, {Dai}, {Dai}, {Dai}, {Danzengluobu}, {Della Volpe}, {Dong}, {Duan}, {Fan}, {Fan}, {Fan}, {Fang}, {Fang}, {Feng}, {Feng}, {Feng}, {Feng}, {Gao}, {Gao}, {Gao}, {Gao}, {Gao}, {Ge}, {Geng}, {Gong}, {Gou}, {Gu}, {Guo}, {Guo}, {Guo}, {Guo}, {Guo}, {Han}, {He}, {He}, {He}, {He}, {He}, {He}, {Heller}, {Hor}, {Hou}, {Hou}, {Hu}, {Hu}, {Hu}, {Hu}, {Huang}, {Huang}, {Huang}, {Huang}, {Huang}, {Huang}, {Ji}, {Ji}, {Jia}, {Jiang}, {Jiang}, {Jin}, {Ke}, {Kuleshov}, {Levochkin}, {Li}, {Li}, {Li}, {Li}, {Li}, {Li}, {Li}, {Li}, {Li}, {Li}, {Li}, {Li}, {Li}, {Li}, {Li}, {Li}, {Li}, {Li}, {Liang}, {Liang}, {Lin}, {Liu}, {Liu}, {Liu}, {Liu}, {Liu}, {Liu}, {Liu}, {Liu},
  {Liu}, {Liu}, {Liu}, {Liu}, {Liu}, {Liu}, {Liu}, {Liu}, {Long}, {Lu}, {Lv}, {Ma}, {Ma}, {Ma}, {Mao}, {Masood}, {Min}, {Mitthumsiri}, {Montaruli}, {Nan}, {Pang}, {Pattarakijwanich}, {Pei}, {Qi}, {Qi}, {Qiao}, {Qin}, {Ruffolo}, {Rulev}, {Saiz}, {Shao}, {Shchegolev}, {Sheng}, {Shi}, {Song}, {Stenkin}, {Stepanov}, {Su}, {Sun}, {Sun}, {Sun}, {Tam}, {Tang}, {Tian}, {Wang}, {Wang}, {Wang}, {Wang}, {Wang}, {Wang}, {Wang}, {Wang}, {Wang}, {Wang}, {Wang}, {Wang}, {Wang}, {Wang}, {Wang}, {Wang}, {Wang}, {Wang}, {Wang}, {Wang}, {Wang}, {Wang}, {Wei}, {Wei}, {Wei}, {Wen}, {Wu}, {Wu}, {Wu}, and {Wu}]{2021Sci...373..425L}
{Lhaaso Collaboration}.; {Cao}, Z.; {Aharonian}, F.; {An}, Q.; {Axikegu}.; {Bai}, L.X.; {Bai}, Y.X.; {Bao}, Y.W.; {Bastieri}, D.; {Bi}, X.J.;  et~al.
\newblock {Peta-electron volt gamma-ray emission from the Crab Nebula}.
\newblock {\em Science} {\bf 2021}, {\em 373},~425--430,  \href{http://arxiv.org/abs/2111.06545}{{\normalfont [arXiv:astro-ph.HE/2111.06545]}}.
\newblock {\url{https://doi.org/10.1126/science.abg5137}}.

\bibitem[{Jokipii}(1966)]{1966ApJ...146..480J}
{Jokipii}, J.R.
\newblock {Cosmic-Ray Propagation. I. Charged Particles in a Random Magnetic Field}.
\newblock {\em \apj} {\bf 1966}, {\em 146},~480.
\newblock {\url{https://doi.org/10.1086/148912}}.

\bibitem[{Jokipii} and {Parker}(1969)]{1969ApJ...155..777J}
{Jokipii}, J.R.; {Parker}, E.N.
\newblock {Stochastic Aspects of Magnetic Lines of Force with Application to Cosmic-Ray Propagation}.
\newblock {\em \apj} {\bf 1969}, {\em 155},~777.
\newblock {\url{https://doi.org/10.1086/149909}}.

\bibitem[{Yan} and {Lazarian}(2002)]{2002PhRvL..89B1102Y}
{Yan}, H.; {Lazarian}, A.
\newblock {Scattering of Cosmic Rays by Magnetohydrodynamic Interstellar Turbulence}.
\newblock {\em \prl} {\bf 2002}, {\em 89},~281102,  \href{http://arxiv.org/abs/astro-ph/0205285}{{\normalfont [arXiv:astro-ph/astro-ph/0205285]}}.
\newblock {\url{https://doi.org/10.1103/PhysRevLett.89.281102}}.

\bibitem[{Qin} et~al.(2002){Qin}, {Matthaeus}, and {Bieber}]{2002ApJ...578L.117Q}
{Qin}, G.; {Matthaeus}, W.H.; {Bieber}, J.W.
\newblock {Perpendicular Transport of Charged Particles in Composite Model Turbulence: Recovery of Diffusion}.
\newblock {\em \apjl} {\bf 2002}, {\em 578},~L117--L120.
\newblock {\url{https://doi.org/10.1086/344687}}.

\bibitem[{Yan} and {Lazarian}(2008)]{2008ApJ...673..942Y}
{Yan}, H.; {Lazarian}, A.
\newblock {Cosmic-Ray Propagation: Nonlinear Diffusion Parallel and Perpendicular to Mean Magnetic Field}.
\newblock {\em \apj} {\bf 2008}, {\em 673},~942--953,  \href{http://arxiv.org/abs/0710.2617}{{\normalfont [arXiv:astro-ph/0710.2617]}}.
\newblock {\url{https://doi.org/10.1086/524771}}.

\bibitem[{Lazarian} and {Xu}(2021)]{2021ApJ...923...53L}
{Lazarian}, A.; {Xu}, S.
\newblock {Diffusion of Cosmic Rays in MHD Turbulence with Magnetic Mirrors}.
\newblock {\em \apj} {\bf 2021}, {\em 923},~53,  \href{http://arxiv.org/abs/2106.08362}{{\normalfont [arXiv:astro-ph.HE/2106.08362]}}.
\newblock {\url{https://doi.org/10.3847/1538-4357/ac2de9}}.

\bibitem[{Kempski} et~al.(2023){Kempski}, {Fielding}, {Quataert}, {Galishnikova}, {Kunz}, {Philippov}, and {Ripperda}]{2023MNRAS.525.4985K}
{Kempski}, P.; {Fielding}, D.B.; {Quataert}, E.; {Galishnikova}, A.K.; {Kunz}, M.W.; {Philippov}, A.A.; {Ripperda}, B.
\newblock {Cosmic ray transport in large-amplitude turbulence with small-scale field reversals}.
\newblock {\em \mnras} {\bf 2023}, {\em 525},~4985--4998,  \href{http://arxiv.org/abs/2304.12335}{{\normalfont [arXiv:astro-ph.HE/2304.12335]}}.
\newblock {\url{https://doi.org/10.1093/mnras/stad2609}}.

\bibitem[{Lemoine}(2023)]{2023JPlPh..89e1701L}
{Lemoine}, M.
\newblock {Particle transport through localized interactions with sharp magnetic field bends in MHD turbulence}.
\newblock {\em Journal of Plasma Physics} {\bf 2023}, {\em 89},~175890501,  \href{http://arxiv.org/abs/2304.03023}{{\normalfont [arXiv:physics.plasm-ph/2304.03023]}}.
\newblock {\url{https://doi.org/10.1017/S0022377823000946}}.

\bibitem[{Ewart} et~al.(2026){Ewart}, {Reichherzer}, {Ren}, {Majeski}, {Mori}, {Nastac}, {Bott}, {Kunz}, and {Schekochihin}]{2026MNRAS.545f2108E}
{Ewart}, R.J.; {Reichherzer}, P.; {Ren}, S.; {Majeski}, S.; {Mori}, F.; {Nastac}, M.L.; {Bott}, A.F.A.; {Kunz}, M.W.; {Schekochihin}, A.A.
\newblock {Cosmic-ray transport in inhomogeneous media}.
\newblock {\em \mnras} {\bf 2026}, {\em 545},~staf2108,  \href{http://arxiv.org/abs/2507.19044}{{\normalfont [arXiv:astro-ph.HE/2507.19044]}}.
\newblock {\url{https://doi.org/10.1093/mnras/staf2108}}.

\bibitem[{Xu} and {Yan}(2013)]{2013ApJ...779..140X}
{Xu}, S.; {Yan}, H.
\newblock {Cosmic-Ray Parallel and Perpendicular Transport in Turbulent Magnetic Fields}.
\newblock {\em \apj} {\bf 2013}, {\em 779},~140,  \href{http://arxiv.org/abs/1307.1346}{{\normalfont [arXiv:astro-ph.HE/1307.1346]}}.
\newblock {\url{https://doi.org/10.1088/0004-637X/779/2/140}}.

\bibitem[{Hu} et~al.(2022){Hu}, {Lazarian}, and {Xu}]{2022MNRAS.512.2111H}
{Hu}, Y.; {Lazarian}, A.; {Xu}, S.
\newblock {Superdiffusion of cosmic rays in compressible magnetized turbulence}.
\newblock {\em \mnras} {\bf 2022}, {\em 512},~2111--2124,  \href{http://arxiv.org/abs/2111.15066}{{\normalfont [arXiv:astro-ph.GA/2111.15066]}}.
\newblock {\url{https://doi.org/10.1093/mnras/stac319}}.

\bibitem[{Lazarian} et~al.(2023){Lazarian}, {Xu}, and {Hu}]{2023FrASS..1054760L}
{Lazarian}, A.; {Xu}, S.; {Hu}, Y.
\newblock {Cosmic ray propagation in turbulent magnetic fields}.
\newblock {\em Frontiers in Astronomy and Space Sciences} {\bf 2023}, {\em 10},~1154760,  \href{http://arxiv.org/abs/2304.02684}{{\normalfont [arXiv:astro-ph.GA/2304.02684]}}.
\newblock {\url{https://doi.org/10.3389/fspas.2023.1154760}}.

\bibitem[{Hu} et~al.(2025){Hu}, {Xu}, {Lazarian}, {Stone}, and {Hopkins}]{2025ApJ...994..142H}
{Hu}, Y.; {Xu}, S.; {Lazarian}, A.; {Stone}, J.M.; {Hopkins}, P.F.
\newblock {Cosmic-ray Perpendicular Superdiffusion and Parallel Mirror Diffusion in a Partially Ionized and Turbulent Medium}.
\newblock {\em \apj} {\bf 2025}, {\em 994},~142,  \href{http://arxiv.org/abs/2505.07421}{{\normalfont [arXiv:astro-ph.GA/2505.07421]}}.
\newblock {\url{https://doi.org/10.3847/1538-4357/ae1127}}.

\bibitem[{Commer{\c{c}}on} et~al.(2019){Commer{\c{c}}on}, {Marcowith}, and {Dubois}]{2019A&A...622A.143C}
{Commer{\c{c}}on}, B.; {Marcowith}, A.; {Dubois}, Y.
\newblock {Cosmic-ray propagation in the bi-stable interstellar medium. I. Conditions for cosmic-ray trapping}.
\newblock {\em \aap} {\bf 2019}, {\em 622},~A143,  \href{http://arxiv.org/abs/1811.11509}{{\normalfont [arXiv:astro-ph.GA/1811.11509]}}.
\newblock {\url{https://doi.org/10.1051/0004-6361/201833809}}.

\bibitem[{Habegger} et~al.(2024){Habegger}, {Ho}, {Yuen}, and {Zweibel}]{2024ApJ...974...17H}
{Habegger}, R.; {Ho}, K.W.; {Yuen}, K.H.; {Zweibel}, E.G.
\newblock {Cosmic-Ray Feedback on Bistable Interstellar Medium Turbulence}.
\newblock {\em \apj} {\bf 2024}, {\em 974},~17,  \href{http://arxiv.org/abs/2403.07976}{{\normalfont [arXiv:astro-ph.HE/2403.07976]}}.
\newblock {\url{https://doi.org/10.3847/1538-4357/ad67da}}.

\bibitem[{Giacinti} et~al.(2012){Giacinti}, {Kachelrie{\ss}}, {Semikoz}, and {Sigl}]{2012JCAP...07..031G}
{Giacinti}, G.; {Kachelrie{\ss}}, M.; {Semikoz}, D.V.; {Sigl}, G.
\newblock {Cosmic ray anisotropy as signature for the transition from galactic to extragalactic cosmic rays}.
\newblock {\em \jcap} {\bf 2012}, {\em 2012},~031,  \href{http://arxiv.org/abs/1112.5599}{{\normalfont [arXiv:astro-ph.HE/1112.5599]}}.
\newblock {\url{https://doi.org/10.1088/1475-7516/2012/07/031}}.

\bibitem[{Snodin} et~al.(2016){Snodin}, {Shukurov}, {Sarson}, {Bushby}, and {Rodrigues}]{2016MNRAS.457.3975S}
{Snodin}, A.P.; {Shukurov}, A.; {Sarson}, G.R.; {Bushby}, P.J.; {Rodrigues}, L.F.S.
\newblock {Global diffusion of cosmic rays in random magnetic fields}.
\newblock {\em \mnras} {\bf 2016}, {\em 457},~3975--3987,  \href{http://arxiv.org/abs/1509.03766}{{\normalfont [arXiv:astro-ph.HE/1509.03766]}}.
\newblock {\url{https://doi.org/10.1093/mnras/stw217}}.

\bibitem[{Kuhlen} et~al.(2025){Kuhlen}, {Mertsch}, and {Phan}]{2025ApJ...992...10K}
{Kuhlen}, M.; {Mertsch}, P.; {Phan}, V.H.M.
\newblock {Diffusion of Relativistic Charged Particles and Field Lines in Isotropic Turbulence. I. Numerical Simulations}.
\newblock {\em \apj} {\bf 2025}, {\em 992},~10,  \href{http://arxiv.org/abs/2211.05881}{{\normalfont [arXiv:astro-ph.HE/2211.05881]}}.
\newblock {\url{https://doi.org/10.3847/1538-4357/adee9a}}.

\bibitem[{Stanimirovic} et~al.(1999){Stanimirovic}, {Staveley-Smith}, {Dickey}, {Sault}, and {Snowden}]{1999MNRAS.302..417S}
{Stanimirovic}, S.; {Staveley-Smith}, L.; {Dickey}, J.M.; {Sault}, R.J.; {Snowden}, S.L.
\newblock {The large-scale HI structure of the Small Magellanic Cloud}.
\newblock {\em \mnras} {\bf 1999}, {\em 302},~417--436.
\newblock {\url{https://doi.org/10.1046/j.1365-8711.1999.02013.x}}.

\bibitem[{V{\'a}zquez-Semadeni} et~al.(2000){V{\'a}zquez-Semadeni}, {Gazol}, and {Scalo}]{2000ApJ...540..271V}
{V{\'a}zquez-Semadeni}, E.; {Gazol}, A.; {Scalo}, J.
\newblock {Is Thermal Instability Significant in Turbulent Galactic Gas?}
\newblock {\em \apj} {\bf 2000}, {\em 540},~271--285,  \href{http://arxiv.org/abs/astro-ph/0001027}{{\normalfont [arXiv:astro-ph/astro-ph/0001027]}}.
\newblock {\url{https://doi.org/10.1086/309318}}.

\bibitem[{Wolfire} et~al.(2003){Wolfire}, {McKee}, {Hollenbach}, and {Tielens}]{2003ApJ...587..278W}
{Wolfire}, M.G.; {McKee}, C.F.; {Hollenbach}, D.; {Tielens}, A.G.G.M.
\newblock {Neutral Atomic Phases of the Interstellar Medium in the Galaxy}.
\newblock {\em \apj} {\bf 2003}, {\em 587},~278--311,  \href{http://arxiv.org/abs/astro-ph/0207098}{{\normalfont [arXiv:astro-ph/astro-ph/0207098]}}.
\newblock {\url{https://doi.org/10.1086/368016}}.

\bibitem[{Kalberla} and {Kerp}(2009)]{2009ARA&A..47...27K}
{Kalberla}, P.M.W.; {Kerp}, J.
\newblock {The Hi Distribution of the Milky Way}.
\newblock {\em \araa} {\bf 2009}, {\em 47},~27--61.
\newblock {\url{https://doi.org/10.1146/annurev-astro-082708-101823}}.

\bibitem[{Draine}(2011)]{2011piim.book.....D}
{Draine}, B.T.
\newblock {\em {Physics of the Interstellar and Intergalactic Medium}};  2011.

\bibitem[{Boulanger} et~al.(2018){Boulanger}, {En{\ss}lin}, {Fletcher}, {Girichides}, {Hackstein}, {Haverkorn}, {H{\"o}randel}, {Jaffe}, {Jasche}, {Kachelrie{\ss}}, {Kotera}, {Pfrommer}, {Rachen}, {Rodrigues}, {Ruiz-Granados}, {Seta}, {Shukurov}, {Sigl}, {Steininger}, {Vacca}, {van der Velden}, {van Vliet}, and {Wang}]{2018JCAP...08..049B}
{Boulanger}, F.; {En{\ss}lin}, T.; {Fletcher}, A.; {Girichides}, P.; {Hackstein}, S.; {Haverkorn}, M.; {H{\"o}randel}, J.R.; {Jaffe}, T.; {Jasche}, J.; {Kachelrie{\ss}}, M.;  et~al.
\newblock {IMAGINE: a comprehensive view of the interstellar medium, Galactic magnetic fields and cosmic rays}.
\newblock {\em \jcap} {\bf 2018}, {\em 2018},~049,  \href{http://arxiv.org/abs/1805.02496}{{\normalfont [arXiv:astro-ph.GA/1805.02496]}}.
\newblock {\url{https://doi.org/10.1088/1475-7516/2018/08/049}}.

\bibitem[{Hu} et~al.(2019){Hu}, {Yuen}, {Lazarian}, {Ho}, {Benjamin}, {Hill}, {Lockman}, {Goldsmith}, and {Lazarian}]{2019NatAs...3..776H}
{Hu}, Y.; {Yuen}, K.H.; {Lazarian}, V.; {Ho}, K.W.; {Benjamin}, R.A.; {Hill}, A.S.; {Lockman}, F.J.; {Goldsmith}, P.F.; {Lazarian}, A.
\newblock {Magnetic field morphology in interstellar clouds with the velocity gradient technique}.
\newblock {\em Nature Astronomy} {\bf 2019}, {\em 3},~776--782,  \href{http://arxiv.org/abs/2002.09948}{{\normalfont [arXiv:astro-ph.GA/2002.09948]}}.
\newblock {\url{https://doi.org/10.1038/s41550-019-0769-0}}.

\bibitem[{Ho} et~al.(2024){Ho}, {Yuen}, and {Lazarian}]{2024arXiv240714199H}
{Ho}, K.W.; {Yuen}, K.H.; {Lazarian}, A.
\newblock {The stable ``Unstable Natural Media'' due to the presence of turbulence}.
\newblock {\em arXiv e-prints} {\bf 2024}, p. arXiv:2407.14199,  \href{http://arxiv.org/abs/2407.14199}{{\normalfont [arXiv:astro-ph.GA/2407.14199]}}.
\newblock {\url{https://doi.org/10.48550/arXiv.2407.14199}}.

\bibitem[{Hu}(2025)]{2025ApJ...986...62H}
{Hu}, Y.
\newblock {Origin of the Multiphase Interstellar Medium: The Effects of Turbulence and Magnetic Field}.
\newblock {\em \apj} {\bf 2025}, {\em 986},~62,  \href{http://arxiv.org/abs/2505.07423}{{\normalfont [arXiv:astro-ph.GA/2505.07423]}}.
\newblock {\url{https://doi.org/10.3847/1538-4357/add731}}.

\bibitem[V{\'a}zquez-Semadeni(2025)]{vazquez2025interstellar}
V{\'a}zquez-Semadeni, E.
\newblock Interstellar Flow and Star Formation {\bf 2025}.

\bibitem[{Hu} et~al.(2026){Hu}, {Truong}, {Hoang}, and {Tram}]{2026arXiv260117255H}
{Hu}, Y.; {Truong}, B.; {Hoang}, T.; {Tram}, L.N.
\newblock {Galactic Dust Polarization in Turbulent Multiphase ISM: On the Origin of the $EE/BB$ Asymmetry}.
\newblock {\em arXiv e-prints} {\bf 2026}, p. arXiv:2601.17255,  \href{http://arxiv.org/abs/2601.17255}{{\normalfont [arXiv:astro-ph.GA/2601.17255]}}.
\newblock {\url{https://doi.org/10.48550/arXiv.2601.17255}}.

\bibitem[{Thomas} et~al.(2025){Thomas}, {Pfrommer}, {Pakmor}, {Lemmerz}, and {Shalaby}]{2025arXiv251016125T}
{Thomas}, T.; {Pfrommer}, C.; {Pakmor}, R.; {Lemmerz}, R.; {Shalaby}, M.
\newblock {Effective cosmic ray diffusion in multiphase galactic environments}.
\newblock {\em arXiv e-prints} {\bf 2025}, p. arXiv:2510.16125,  \href{http://arxiv.org/abs/2510.16125}{{\normalfont [arXiv:astro-ph.GA/2510.16125]}}.
\newblock {\url{https://doi.org/10.48550/arXiv.2510.16125}}.

\bibitem[{Stone} et~al.(2024){Stone}, {Mullen}, {Fielding}, {Grete}, {Guo}, {Kempski}, {Most}, {White}, and {Wong}]{2024arXiv240916053S}
{Stone}, J.M.; {Mullen}, P.D.; {Fielding}, D.; {Grete}, P.; {Guo}, M.; {Kempski}, P.; {Most}, E.R.; {White}, C.J.; {Wong}, G.N.
\newblock {AthenaK: A Performance-Portable Version of the Athena++ AMR Framework}.
\newblock {\em arXiv e-prints} {\bf 2024}, p. arXiv:2409.16053,  \href{http://arxiv.org/abs/2409.16053}{{\normalfont [arXiv:astro-ph.IM/2409.16053]}}.
\newblock {\url{https://doi.org/10.48550/arXiv.2409.16053}}.

\bibitem[{Koyama} and {Inutsuka}(2002)]{2002ApJ...564L..97K}
{Koyama}, H.; {Inutsuka}, S.i.
\newblock {An Origin of Supersonic Motions in Interstellar Clouds}.
\newblock {\em \apjl} {\bf 2002}, {\em 564},~L97--L100,  \href{http://arxiv.org/abs/astro-ph/0112420}{{\normalfont [arXiv:astro-ph/astro-ph/0112420]}}.
\newblock {\url{https://doi.org/10.1086/338978}}.

\bibitem[{Crutcher}(2012)]{2012ARA&A..50...29C}
{Crutcher}, R.M.
\newblock {Magnetic Fields in Molecular Clouds}.
\newblock {\em \araa} {\bf 2012}, {\em 50},~29--63.
\newblock {\url{https://doi.org/10.1146/annurev-astro-081811-125514}}.

\bibitem[{Larson}(1981)]{1981MNRAS.194..809L}
{Larson}, R.B.
\newblock {Turbulence and star formation in molecular clouds.}
\newblock {\em \mnras} {\bf 1981}, {\em 194},~809--826.
\newblock {\url{https://doi.org/10.1093/mnras/194.4.809}}.

\bibitem[{Ha} et~al.(2022){Ha}, {Li}, {Kounkel}, {Xu}, {Li}, and {Zheng}]{2022ApJ...934....7H}
{Ha}, T.; {Li}, Y.; {Kounkel}, M.; {Xu}, S.; {Li}, H.; {Zheng}, Y.
\newblock {Turbulence in Milky Way Star-forming Regions Traced by Young Stars and Gas}.
\newblock {\em \apj} {\bf 2022}, {\em 934},~7,  \href{http://arxiv.org/abs/2205.00012}{{\normalfont [arXiv:astro-ph.GA/2205.00012]}}.
\newblock {\url{https://doi.org/10.3847/1538-4357/ac76bf}}.

\bibitem[{Hu} et~al.(2024){Hu}, {Xu}, {Arzamasskiy}, {Stone}, and {Lazarian}]{2024MNRAS.527.3945H}
{Hu}, Y.; {Xu}, S.; {Arzamasskiy}, L.; {Stone}, J.M.; {Lazarian}, A.
\newblock {Damping of MHD turbulence in a partially ionized medium}.
\newblock {\em \mnras} {\bf 2024}, {\em 527},~3945--3961,  \href{http://arxiv.org/abs/2306.10010}{{\normalfont [arXiv:astro-ph.GA/2306.10010]}}.
\newblock {\url{https://doi.org/10.1093/mnras/stad3493}}.

\bibitem[{Lazarian} and {Yan}(2014)]{2014ApJ...784...38L}
{Lazarian}, A.; {Yan}, H.
\newblock {Superdiffusion of Cosmic Rays: Implications for Cosmic Ray Acceleration}.
\newblock {\em \apj} {\bf 2014}, {\em 784},~38,  \href{http://arxiv.org/abs/1308.3244}{{\normalfont [arXiv:astro-ph.HE/1308.3244]}}.
\newblock {\url{https://doi.org/10.1088/0004-637X/784/1/38}}.

\bibitem[{Aartsen} et~al.(2013){Aartsen}, {Abbasi}, {Abdou}, {Ackermann}, {Adams}, {Aguilar}, {Ahlers}, {Altmann}, {Andeen}, {Auffenberg}, {Bai}, {Baker}, {Barwick}, {Baum}, {Bay}, {Beattie}, {Beatty}, {Bechet}, {Becker Tjus}, {Becker}, {Bell}, {Benabderrahmane}, {BenZvi}, {Berdermann}, {Berghaus}, {Berley}, {Bernardini}, {Bertrand}, {Besson}, {Bindig}, {Bissok}, {Blaufuss}, {Blumenthal}, {Boersma}, {Bohaichuk}, {Bohm}, {Bose}, {B{\"o}ser}, {Botner}, {Brayeur}, {Brown}, {Bruijn}, {Brunner}, {Carson}, {Casey}, {Casier}, {Chirkin}, {Christy}, {Clark}, {Clevermann}, {Cohen}, {Cowen}, {Cruz Silva}, {Danninger}, {Daughhetee}, {Davis}, {De Clercq}, {De Ridder}, {Descamps}, {Desiati}, {de Vries-Uiterweerd}, {DeYoung}, {D{\'\i}az-V{\'e}lez}, {Dreyer}, {Dumm}, {Dunkman}, {Eagan}, {Eisch}, {Ellsworth}, {Engdeg{\r{a}}rd}, {Euler}, {Evenson}, {Fadiran}, {Fazely}, {Fedynitch}, {Feintzeig}, {Feusels}, {Filimonov}, {Finley}, {Fischer-Wasels}, {Flis}, {Franckowiak}, {Franke}, {Frantzen}, {Fuchs}, {Gaisser}, {Gallagher},
  {Gerhardt}, {Gladstone}, {Gl{\"u}senkamp}, {Goldschmidt}, {Golup}, {Goodman}, {G{\'o}ra}, {Grant}, {Gross}, {Grullon}, {Gurtner}, {Ha}, {Haj Ismail}, {Hallgren}, {Halzen}, {Hanson}, {Heereman}, {Heimann}, {Heinen}, {Helbing}, {Hellauer}, {Hickford}, {Hill}, {Hoffman}, {Hoffmann}, {Homeier}, {Hoshina}, {Huelsnitz}, {Hulth}, {Hultqvist}, {Hussain}, {Ishihara}, {Jacobi}, {Jacobsen}, {Japaridze}, {Jlelati}, {Kappes}, {Karg}, {Karle}, {Kiryluk}, {Kislat}, {Kl{\"a}s}, {Klein}, {K{\"o}hne}, {Kohnen}, {Kolanoski}, {K{\"o}pke}, {Kopper}, {Kopper}, {Koskinen}, {Kowalski}, {Krasberg}, {Kroll}, {Kunnen}, {Kurahashi}, {Kuwabara}, {Labare}, {Landsman}, {Larson}, {Lauer}, {Lesiak-Bzdak}, {L{\"u}nemann}, {Madsen}, {Maruyama}, {Mase}, {Matis}, {McNally}, {Meagher}, {Merck}, {M{\'e}sz{\'a}ros}, {Meures}, {Miarecki}, {Middell}, {Milke}, {Miller}, {Mohrmann}, {Montaruli}, {Morse}, {Nahnhauer}, {Naumann}, {Nowicki}, {Nygren}, {Obertacke}, {Odrowski}, {Olivas}, {Olivo}, {O'Murchadha}, {Panknin}, {Paul}, {Pepper}, {P{\'e}rez de
  los Heros}, {Pieloth}, {Pirk}, {Posselt}, {Price}, {Przybylski}, {R{\"a}del}, {Rawlins}, {Redl}, {Resconi}, {Rhode}, {Ribordy}, {Richman}, {Riedel}, {Rodrigues}, {Rothmaier}, {Rott}, {Ruhe}, {Ruzybayev}, {Ryckbosch}, {Saba}, {Salameh}, and {Sander}]{2013ApJ...765...55A}
{Aartsen}, M.G.; {Abbasi}, R.; {Abdou}, Y.; {Ackermann}, M.; {Adams}, J.; {Aguilar}, J.A.; {Ahlers}, M.; {Altmann}, D.; {Andeen}, K.; {Auffenberg}, J.;  et~al.
\newblock {Observation of Cosmic-Ray Anisotropy with the IceTop Air Shower Array}.
\newblock {\em \apj} {\bf 2013}, {\em 765},~55,  \href{http://arxiv.org/abs/1210.5278}{{\normalfont [arXiv:astro-ph.HE/1210.5278]}}.
\newblock {\url{https://doi.org/10.1088/0004-637X/765/1/55}}.

\bibitem[{Aartsen} et~al.(2016){Aartsen}, {Abraham}, {Ackermann}, {Adams}, {Aguilar}, {Ahlers}, {Ahrens}, {Altmann}, {Anderson}, {Ansseau}, {Anton}, {Archinger}, {Arguelles}, {Arlen}, {Auffenberg}, {Bai}, {Barwick}, {Baum}, {Bay}, {Beatty}, {Becker Tjus}, {Becker}, {Beiser}, {BenZvi}, {Berghaus}, {Berley}, {Bernardini}, {Bernhard}, {Besson}, {Binder}, {Bindig}, {Bissok}, {Blaufuss}, {Blumenthal}, {Boersma}, {Bohm}, {B{\"o}rner}, {Bos}, {Bose}, {B{\"o}ser}, {Botner}, {Braun}, {Brayeur}, {Bretz}, {Buzinsky}, {Casey}, {Casier}, {Cheung}, {Chirkin}, {Christov}, {Clark}, {Classen}, {Coenders}, {Collin}, {Conrad}, {Cowen}, {Cruz Silva}, {Daughhetee}, {Davis}, {Day}, {de Andr{\'e}}, {De Clercq}, {del Pino Rosendo}, {Dembinski}, {De Ridder}, {Desiati}, {de Vries}, {de Wasseige}, {de With}, {DeYoung}, {D{\'\i}az-V{\'e}lez}, {di Lorenzo}, {Dujmovic}, {Dumm}, {Dunkman}, {Eberhardt}, {Ehrhardt}, {Eichmann}, {Euler}, {Evenson}, {Fahey}, {Fazely}, {Feintzeig}, {Felde}, {Filimonov}, {Finley}, {Flis}, {F{\"o}sig}, {Fuchs},
  {Gaisser}, {Gaior}, {Gallagher}, {Gerhardt}, {Ghorbani}, {Gier}, {Gladstone}, {Glagla}, {Gl{\"u}senkamp}, {Goldschmidt}, {Golup}, {Gonzalez}, {G{\'o}ra}, {Grant}, {Griffith}, {Ha}, {Haack}, {Haj Ismail}, {Hallgren}, {Halzen}, {Hansen}, {Hansmann}, {Hansmann}, {Hanson}, {Hebecker}, {Heereman}, {Helbing}, {Hellauer}, {Hickford}, {Hignight}, {Hill}, {Hoffman}, {Hoffmann}, {Holzapfel}, {Homeier}, {Hoshina}, {Huang}, {Huber}, {Huelsnitz}, {Hulth}, {Hultqvist}, {In}, {Ishihara}, {Jacobi}, {Japaridze}, {Jeong}, {Jero}, {Jones}, {Jurkovic}, {Kappes}, {Karg}, {Karle}, {Katz}, {Kauer}, {Keivani}, {Kelley}, {Kemp}, {Kheirandish}, {Kim}, {Kintscher}, {Kiryluk}, {Klein}, {Kohnen}, {Koirala}, {Kolanoski}, {Konietz}, {K{\"o}pke}, {Kopper}, {Kopper}, {Koskinen}, {Kowalski}, {Krings}, {Kroll}, {Kroll}, {Kr{\"u}ckl}, {Kunnen}, {Kunwar}, {Kurahashi}, {Kuwabara}, {Labare}, {Lanfranchi}, {Larson}, {Lennarz}, {Lesiak-Bzdak}, {Leuermann}, {Leuner}, {Lu}, {L{\"u}nemann}, {Madsen}, {Maggi}, {Mahn}, {Mandelartz}, {Maruyama}, {Mase},
  {Matis}, {Maunu}, {McNally}, {Meagher}, {Medici}, {Meier}, {Meli}, {Menne}, {Merino}, {Meures}, {Miarecki}, {Middell}, {Mohrmann}, {Montaruli}, {Morse}, {Nahnhauer}, and {Naumann}]{2016ApJ...826..220A}
{Aartsen}, M.G.; {Abraham}, K.; {Ackermann}, M.; {Adams}, J.; {Aguilar}, J.A.; {Ahlers}, M.; {Ahrens}, M.; {Altmann}, D.; {Anderson}, T.; {Ansseau}, I.;  et~al.
\newblock {Anisotropy in Cosmic-Ray Arrival Directions in the Southern Hemisphere Based on Six Years of Data from the IceCube Detector}.
\newblock {\em \apj} {\bf 2016}, {\em 826},~220,  \href{http://arxiv.org/abs/1603.01227}{{\normalfont [arXiv:astro-ph.HE/1603.01227]}}.
\newblock {\url{https://doi.org/10.3847/0004-637X/826/2/220}}.

\bibitem[{Amenomori} et~al.(2006){Amenomori}, {Ayabe}, {Bi}, {Chen}, {Cui}, {Danzengluobu}, {Ding}, {Ding}, {Feng}, {Feng}, {Feng}, {Gao}, {Geng}, {Guo}, {He}, {He}, {Hibino}, {Hotta}, {Hu}, {Hu}, {Huang}, {Huang}, {Jia}, {Kajino}, {Kasahara}, {Katayose}, {Kato}, {Kawata}, {Labaciren}, {Le}, {Li}, {Li}, {Lou}, {Lu}, {Lu}, {Meng}, {Mizutani}, {Mu}, {Munakata}, {Nagai}, {Nanjo}, {Nishizawa}, {Ohnishi}, {Ohta}, {Onuma}, {Ouchi}, {Ozawa}, {Ren}, {Saito}, {Saito}, {Sakata}, {Sako}, {Sasaki}, {Shibata}, {Shiomi}, {Shirai}, {Sugimoto}, {Takita}, {Tan}, {Tateyama}, {Torii}, {Tsuchiya}, {Udo}, {Wang}, {Wang}, {Wang}, {Wang}, {Wu}, {Xue}, {Yamamoto}, {Yan}, {Yang}, {Yasue}, {Ye}, {Yu}, {Yuan}, {Yuda}, {Zhang}, {Zhang}, {Zhang}, {Zhang}, {Zhang}, {Zhang}, {Zhaxisangzhu}, {Zhou}, and {Tibet AS{\ensuremath{\gamma}} Collaboration}]{2006Sci...314..439A}
{Amenomori}, M.; {Ayabe}, S.; {Bi}, X.J.; {Chen}, D.; {Cui}, S.W.; {Danzengluobu}.; {Ding}, L.K.; {Ding}, X.H.; {Feng}, C.F.; {Feng}, Z.;  et~al.
\newblock {Anisotropy and Corotation of Galactic Cosmic Rays}.
\newblock {\em Science} {\bf 2006}, {\em 314},~439--443,  \href{http://arxiv.org/abs/astro-ph/0610671}{{\normalfont [arXiv:astro-ph/astro-ph/0610671]}}.
\newblock {\url{https://doi.org/10.1126/science.1131702}}.

\bibitem[{Liu} et~al.(2019){Liu}, {Yan}, and {Zhang}]{2019PhRvL.123v1103L}
{Liu}, R.Y.; {Yan}, H.; {Zhang}, H.
\newblock {Understanding the Multiwavelength Observation of Geminga's Tev Halo: The Role of Anisotropic Diffusion of Particles}.
\newblock {\em \prl} {\bf 2019}, {\em 123},~221103,  \href{http://arxiv.org/abs/1904.11536}{{\normalfont [arXiv:astro-ph.HE/1904.11536]}}.
\newblock {\url{https://doi.org/10.1103/PhysRevLett.123.221103}}.

\bibitem[{De La Torre Luque} et~al.(2022){De La Torre Luque}, {Fornieri}, and {Linden}]{2022PhRvD.106l3033D}
{De La Torre Luque}, P.; {Fornieri}, O.; {Linden}, T.
\newblock {Anisotropic diffusion cannot explain TeV halo observations}.
\newblock {\em \prd} {\bf 2022}, {\em 106},~123033,  \href{http://arxiv.org/abs/2205.08544}{{\normalfont [arXiv:astro-ph.HE/2205.08544]}}.
\newblock {\url{https://doi.org/10.1103/PhysRevD.106.123033}}.

\bibitem[{Kulsrud} and {Cesarsky}(1971)]{1971ApL.....8..189K}
{Kulsrud}, R.M.; {Cesarsky}, C.J.
\newblock {The Effectiveness of Instabilities for the Confinement of High Energy Cosmic Rays in the Galactic Disk}.
\newblock {\em \apjl} {\bf 1971}, {\em 8},~189.

\bibitem[{Xu} et~al.(2016){Xu}, {Yan}, and {Lazarian}]{2016ApJ...826..166X}
{Xu}, S.; {Yan}, H.; {Lazarian}, A.
\newblock {Damping of Magnetohydrodynamic Turbulence in Partially Ionized Plasma: Implications for Cosmic Ray Propagation}.
\newblock {\em \apj} {\bf 2016}, {\em 826},~166,  \href{http://arxiv.org/abs/1506.05585}{{\normalfont [arXiv:astro-ph.HE/1506.05585]}}.
\newblock {\url{https://doi.org/10.3847/0004-637X/826/2/166}}.

\bibitem[{Plotnikov} et~al.(2021){Plotnikov}, {Ostriker}, and {Bai}]{2021ApJ...914....3P}
{Plotnikov}, I.; {Ostriker}, E.C.; {Bai}, X.N.
\newblock {Influence of Ion-Neutral Damping on the Cosmic-Ray Streaming Instability: Magnetohydrodynamic Particle-in-cell Simulations}.
\newblock {\em \apj} {\bf 2021}, {\em 914},~3,  \href{http://arxiv.org/abs/2102.11878}{{\normalfont [arXiv:astro-ph.HE/2102.11878]}}.
\newblock {\url{https://doi.org/10.3847/1538-4357/abf7b3}}.

\bibitem[{Lazarian} and {Xu}(2022)]{2022FrP....10.2799L}
{Lazarian}, A.; {Xu}, S.
\newblock {Damping of Alfv{\'e}n Waves in MHD Turbulence and Implications for Cosmic Ray Streaming Instability and Galactic Winds}.
\newblock {\em Frontiers in Physics} {\bf 2022}, {\em 10},~702799,  \href{http://arxiv.org/abs/2201.05168}{{\normalfont [arXiv:astro-ph.GA/2201.05168]}}.
\newblock {\url{https://doi.org/10.3389/fphy.2022.702799}}.

\end{thebibliography}

\PublishersNote{}
\end{adjustwidth}
\end{document}